\documentclass{JHEP3}
\usepackage{epsfig}
\usepackage{graphicx}
\usepackage{dcolumn}
\usepackage{bm}

\def\simleq{\; \raise0.3ex\hbox{$<$\kern-0.75em
      \raise-1.1ex\hbox{$\sim$}}\; }
\def\simgeq{\; \raise0.3ex\hbox{$>$\kern-0.75em
      \raise-1.1ex\hbox{$\sim$}}\; }

\def\bea{\begin{eqnarray}}
\def\eea{\end{eqnarray}}
\def\be{\begin{eqnarray}}
\def\ee{\end{eqnarray}}
 
\def\lesssim{\mathrel{\hbox{\rlap{\hbox{\lower4pt\hbox{$\sim$}}}\hbox{$<$}}}}
\def\gtrsim{\mathrel{\hbox{\rlap{\hbox{\lower4pt\hbox{$\sim$}}}\hbox{$>$}}}}
\newcommand{\ft}[2]{{\textstyle\frac{#1}{#2}}}

\def\rme{{\rm e}}
\def\rmi{{\rm i}}
\def\rmd{{\rm d}}
\newsavebox{\uuunit}
\sbox{\uuunit}
    {\setlength{\unitlength}{0.825em}
     \begin{picture}(0.6,0.7)
        \thinlines
        \put(0,0){\line(1,0){0.5}}
        \put(0.15,0){\line(0,1){0.7}}
        \put(0.35,0){\line(0,1){0.8}}
       \multiput(0.3,0.8)(-0.04,-0.02){12}{\rule{0.5pt}{0.5pt}}
     \end {picture}}

\csname @addtoreset\endcsname{equation}{section}

\newcommand{\N}{\mathop{\mathcal{N}}}

\title{D3/D7 Brane Inflation  and Semilocal Strings}

\author{Keshav Dasgupta, Jonathan P. Hsu, Renata Kallosh, Andrei Linde,
 Marco Zagermann
\\

Department of Physics, Stanford University,
\\ Stanford, CA 94305-4060, USA.\\
e-mail: keshav@itp.stanford.edu, pihsu@stanford.edu,
kallosh@stanford.edu,  alinde@stanford.edu,
zagermann@itp.stanford.edu}

\received{\today}       
 \preprint{SU-ITP-04/21\\  ~hep-th/0405247\\ May 26 2004}

\abstract{\small Among the inflationary models based on string theory, the D3/D7 model has the advantage that the flatness of the inflaton potential can be protected even with moduli
stabilization. However, the Abrikosov-Nielsen-Olesen BPS cosmic strings produced at the end of original D3/D7 inflation lead to an additional contribution to the CMB anisotropy. To make this
contribution consistent with the WMAP results one needs an extremely small gauge coupling in
the effective D-term inflation model. Such couplings may be difficult to justify in string theory. Here we develop a generalized version of the D3/D7 brane model, which leads to
semilocal strings, instead of the topologically stable ANO cosmic strings. We show that the semilocal strings have unbroken supersymmetry when embedded into supergravity with FI terms. The energy of these strings is independent of their thickness. We confirm the existing arguments that strings of such type disappear soon after their formation and do not pose any cosmological problems, for any value of the gauge coupling. This should simplify the task of constructing fully realistic models of D3/D7 inflation.  }

\begin{document}

\section{Introduction}

Despite the recent period of intense research activity on inflationary models in string theory, it is still a challenging task to
derive a realistic model of inflation  consistent
with the stabilization of all string moduli.
A particularly important problem is to reconcile the original idea of
brane inflation \cite{Dvali:1998pa,DHHK}
with an appropriate mechanism of volume stabilization so as to make it
a valid model of
string theory. One of the most developed models in this context
was suggested in \cite{KKLMMT} based on the
mechanisms of  volume stabilization proposed in
\cite{KKLT,Burgess:2003ic}. Recently a particular class of models in string theory
with moduli stabilization was constructed in \cite{Denef:2004dm}. A major problem associated with
inflationary models in  warped compactifications of
type IIB string theory has to do with the fact that the inflaton
in these models (the position of the D3 brane deep inside the warped
throat geometry) is a conformally coupled scalar in the effective
four-dimensional near de Sitter geometry. Therefore, the mass of
such a scalar, $m^2_\phi$, is close to $2H^2= {1\over 6} R$, where
$H^2$ is the Hubble parameter and $R$ is the curvature of de
Sitter space. This does not meet  the observational requirement $m^2_\phi\sim 10^{-2} H^2$,
as was uncovered in \cite{KKLMMT}, where some possible resolutions of this problem were suggested. Additional
attempts to solve this problem in closely related settings were
recently proposed in
\cite{DeWolfe:2004qx}. New ideas for string
cosmology based on non-linear Dirac-Born-Infeld brane actions were
developed in \cite{Silverstein:2003hf}.

On the other hand, it has been found recently
\cite{Hsu:2004hi,KalloshHsuProk,KTW,Kallosh:2004rs} that a particular brane
inflation model, the  D3/D7 model \cite{DHHK}, features a shift
symmetry for the inflaton field. The inflaton in this model is related to the
position of a D3 brane with respect to a D7 brane in a $K3\times
{T^2\over \mathbb{Z}_2}$ compactification of type IIB string
theory, where  $\mathbb{Z}_2$ involves both orbifold as well as
orientifold operations, $\mathbb{Z}_{2}= \mathcal{I}_{2}\cdot
\Omega \cdot (-1)^{F_{L}}$. \footnote{We use the standard notation
$\mathcal{I}_{2}$ and $\Omega$ to denote, respectively, the
orbifold and orientifold action, whereas $(-1)^{F_{L}}$ reverses
the left-moving spacetime fermion number.} The inflaton shift
symmetry follows from  the special geometry in such models, as described
in ref. \cite{Angelantonj:2003zx}.

The D3/D7 model \cite{DHHK} has an effective description as a
hybrid inflation \cite{L}, or, more specifically,  as a D-term inflation
model \cite{Binetruy:1996xj,BDKV}. The  flat
direction of the inflaton potential is associated with the shift
symmetry, which is preserved after the stabilization of the volume modulus.
It is broken spontaneously only at the quantum level.  This
symmetry protects the small inflaton mass during inflation.   This
is an important advantage, since in other models of inflation
in string theory it is difficult to protect the inflaton
field in the early universe from acquiring an unacceptably large
mass.

In its original form, however, the D3/D7 model \cite{DHHK} with $U(1)$ gauge
symmetry has another problem. At the end of inflation, cosmic strings form,
giving a contribution to the primordial density perturbations \cite{pterm}.
Such stringy perturbations do not lead to the multiple peak structure in the
spectrum of the CMB anisotropies. The data from WMAP show that the contribution
of the cosmic strings to the CMB anisotropy should not exceed a few percent of
the standard inflationary perturbations
\cite{Avelino:2003nn,Pogosian:2004mi}. In a general class of  brane
inflation models, the production, spectrum and the evolution of cosmic strings
were studied in
\cite{Kofman:2002rh,Jones:2003da,Dvali:2003zj,Dvali:2003zh,Halyo:2003uu,BDKV}.

It is interesting to note that the model \cite{KKLMMT} of $D3 /
{\overline{D3}}$ inflation in the highly warped throat has a natural mechanism for suppressing the
contribution of the cosmic strings to the CMB anisotropy
\cite{KKLMMT}: The effective tension of the produced strings and,
therefore, their contribution to metric perturbations,  is
exponentially suppressed due to the large warp factor. The
production of
very light cosmic strings in this model and in its generalizations
was recently studied in \cite{Copeland:2003bj}.
Note that in the warped region of the geometry there is an $m^2 \sim H^2$ 
problem with inflation. On the other hand, large warping is known to relieve the 
problem with cosmic strings. 
In the D3/D7
model, by contrast, we do not have the  $m^2\sim H^2$ problem,
but we also do not have an automatic solution of the cosmic string
problem.

There are many ways to address this problem.  One may change the theory in such
a way as to make the potential somewhat asymmetric, so that stable strings are
not formed \cite{Lazar,Lyth:1998xn}.  Alternatively, one can consider models
with extremely small coupling constants, which suppresses the stringy
contribution \cite{K01,pterm,Endo:2003fr}. In particular it was shown in
\cite{pterm} that if one assumes, following \cite{Avelino:2003nn}, that the
contribution of cosmic strings to the amplitude of metric perturbations should
be smaller than $O(10^{-2})$ of the inflationary contribution, then the
effective $g^2$ must be smaller than $10^{-10}$. Meanwhile if one follows
recent work \cite{Pogosian:2004mi} and assumes that the stringy contribution
should not exceed $O(10^{-1})$ of the inflationary contribution, then the
constraint on $g^2$ which can be obtained following Ref. \cite{pterm} becomes
somewhat less stringent, but it still remains quite strong: \, $g^2 \lesssim
10^{-7}$. In the context of the D3/D7 model with volume stabilization, one
expects that the effective $g^2$ comes from the D3 brane gauge coupling. It may
then be rather difficult to get $g^2 \lesssim 10^{-7}$ in this context.

A simple and elegant method for solving the cosmic string problem
in D-term inflation models  has recently been proposed in
\cite{UAD} and \cite{BDKV}. If one simply replicates the charged
matter multiplets responsible for the string formation so as to
obtain a non-trivial global symmetry such as $SU(2)$, one can
render the vacuum manifold simply connected, and the relevant
cosmic strings become so-called  \emph{semilocal} strings
\cite{VA1}-\cite{Edelsteinsemilocal}. The stability properties of semilocal
strings cannot be inferred from topological considerations alone,
but also depend on the values of the couplings in the underlying
Lagrangian \cite{VA1}-\cite{Preskill}.
 Numerical simulations
of their evolution suggest,   that they do not pose any cosmological
problems \cite{ABL99}. As an alternative way to circumvent the cosmic
 string problem,
 it was suggested in \cite{BDKV}
to embed the D-term inflation model into some model with larger
\emph{gauge} symmetry so that the strings would not be
topologically stable.

Following these lines, we will, in this paper,  generalize the original version
of the $D3/D7$ inflationary model \cite{DHHK}
in such a way as to include a larger set of global and/or local
symmetries. In order to do so, we first go back to the original model
\cite{DHHK} and discuss the precise relation
between the  brane construction and the associated four dimensional ${\cal N}=2$
supersymmetric gauge theory.
In this discussion, we
will also carefully take into
account the volume modulus dependence at each step so as  to update the
model with regard  to the more recent work on stabilization.
The purpose of revisiting the original model \cite{DHHK}
  is to show how exactly
the surviving interacting $U(1)$ gauge group of the effective
D-term inflation model   is related to the two $U(1)$'s from
the D3 and D7 worldvolume theories. This will give us a much
better understanding of the original model and will allow us to
proceed with generalized models with higher symmetries resulting
from   brane systems with multiple branes.
Furthermore, when we embed this discussion
   into the compact model of \cite{DHHK} but now with
stacks of coincident  D3 or D7 branes, we will show that the underlying
F-theory picture correctly predicts the local and global symmetries
one expects for the cosmology of the model.

Fluxes and non-perturbative effects, providing stabilization of
moduli in string theory \cite{Giddings:2001yu, KKLT}, as well as
embedding of the relevant 4-dimensional gauge theory with FI term
into 4-dimensional supergravity \cite{K01}, break ${\cal N}=2$
supersymmetry to ${\cal N}=1$ supersymmetry. The embedding of the
standard
Abrikosov-Nielsen-Olesen (ANO) cosmic strings
into ${\cal N}=1$ supergravity with FI terms have been studied recently
in  \cite{Dvali:2003zh}.
They were shown to saturate the BPS bound and to preserve
 one half of the original $\mathcal{N}=1$ local supersymmetry
  \cite{Dvali:2003zh}. Their
possible relation to D1 strings and wrapped D3 branes in string
theory was discussed in \cite{Dvali:2003zh,Halyo:2003uu,BDKV}.

As we  show in this paper, the semilocal strings relevant for our
generalized D3/D7 inflationary model with global symmetries can likewise be realized  as 1/2-BPS solutions in $\mathcal{N}=1$ supergravity.
More precisely, we show that there are families of 1/2-supersymmetric
semilocal string solutions parameterized by an undetermined integration
constant that controls the thickness (or the ``spreading'') of the flux
tube in the string core.
These findings are in agreement with earlier results in
related non-supersymmetric models \cite{Hindmarsh1,Hindmarsh2,GORS}
and imply that it does not cost any energy for
  the semilocal strings of our model to change their thickness.

Finally, we will discuss the production and the cosmological evolution of semilocal strings after
inflation in our scenario. Unlike the topologically stable ANO strings, which form loops or can be infinitely long, the semilocal strings formed after inflation look like a collection of loops and open string segments of finite length with monopoles at their ends. This fact, together with the possibility of their spreading at no energy cost, may lead to  the absence of long semilocal strings, and, as a result, to the absence of the string-related  large-scale  metric perturbations.   Indeed, numerical simulations suggest that semilocal strings disappear soon after their formation
\cite{ABL99}. However, these simulations were performed in  the context of a slightly different model, without
taking into account the cosmological evolution. For a full quantum
mechanical investigation of this issue one can use the methods
of lattice simulations of symmetry breaking after hybrid inflation
\cite{Felder:2000hj}. We hope to perform these calculations in a
separate publication. In this paper we restrict ourselves to a simple
qualitative analysis of the situation. We will argue that the
main conclusions of Ref. \cite{ABL99} should be valid for the
semilocal strings produced after inflation in the model studied in
\cite{UAD} and in our paper. This suggests that the generalized
D3/D7 model developed in our paper does not have any problems with
cosmological perturbations produced by strings, and therefore
there is no need to require that the coupling constant $g^2$ must be very
small.

Thus,  the purpose of this paper is
threefold: i) To develop the generalized D3/D7 brane model
including the volume modulus as well as some global symmetries in
addition to the local $U(1)$ gauge symmetry; such models will have
semilocal cosmic strings instead of ANO strings. ii) To establish
the unbroken supersymmetry of these semilocal strings embedded
into ${\cal N}=1$ supergravity with FI terms for arbitrary
``spread'' parameter. iii) To critically analyse and confirm the
arguments existing in the literature \cite{UAD}  that the models
with semilocal strings have no cosmological problems.

The organization of this paper is as follows: In section 2 we take
a single D3 and a single D7 that is wrapped on a $K3$ manifold
with non-compact orthogonal directions. We study the precise world
volume actions of the branes when the system is kept in a curved
background with appropriate fluxes turned on. This analysis will
serve as a dictionary between the world volume coordinates and the
known P-term action studied earlier in \cite{pterm}. In section 2
we do not take the back-reactions of branes and fluxes on the
geometry into account. This is addressed in section 3 where we
also consider a fully compact scenario using the underlying
F-theory picture that was developed in \cite{DHHK}. We show that
the F-theory curve correctly reproduces the local and the global
symmetries for the system when we have multiple D3 and D7 branes.
The back-reactions of branes and fluxes are shown to be given by
warp factors that are typically of order one when the $K3$
manifold is fixed at a large volume. In section 4 we come to the
analysis of semi-local strings in this model with $U(1)$ gauge
symmetry and a global $SU(2)$ symmetry. Earlier studies of
semi-local strings in the literature did not address the precise
connection with local ${\cal N} = 1$ supersymmetry. We show that
the semi-local strings can arise as 1/2 BPS solitons in the
extended Abelian Higgs model embedded in ${\cal N} = 1$
supergravity.  Finally in section 5 we discuss the cosmology of
the model. We discuss the issue of formation of semi-local strings
and their effect on the large scale density perturbations of the
universe. We argue that these strings do not cause any problems
with cosmological perturbations at any coupling. Therefore,
 we are not restricted to keep the
coupling constant $g^2$ very small. This may  help to
 construct fully realistic models of D3/D7 inflation.

\section{The $D3/D7$ model, P-term model and the volume modulus}

In the first version  of the $D3/D7$ model of inflation \cite{DHHK}
the internal manifold was assumed to be compact, but the mechanism
of stabilization of the volume modulus was not discussed. However,
later in \cite{KalloshHsuProk,Hsu:2004hi,Kallosh:2004rs} the
volume modulus and its stabilization in the ``inflaton trench''
were included into the effective D-term inflation model. Here we
will  study the dependence on the volume modulus in the $D3/D7$
brane system by carefully keeping track of the powers of the
volume, and we will also clarify the origin of the effective
4-dimensional model. Postponing the discussion of six compact dimensions to section 3, we will first   consider   putting the $D3/D7$ system on
a compact K3 with the $D7$ wrapped on the K3. We will also give
  an explicit map between the
 usual D-brane variables and the variables
used in cosmology from $d=4$ supergravity as in \cite{pterm}.

The basic features of the D3/D7 inflationary model \cite{DHHK} can
be easily understood  from the general properties of $Dp/D(p+4)$
brane systems (see e.g. \cite{pbook}). As worked out in
\cite{pbook} for the special case $p=5$, $5-9$ strings have charge
$+1$ with respect to the $U(1)$ gauge field on the 9-brane and
charge $-1$ with respect to the $U(1)$ gauge field on the 5-brane,
whereas for $9-5$ strings the signs are reversed due to their
opposite orientation. Two T-dualities then yield the system with
$p=3$ which we are interested in. We thus have a D3/D7 system with
$3-7$ and $7-3$ strings stretched between the D3 and D7 branes
with completely analogous charge assignments.

In the case when the fluxes of the D7 gauge field in the
directions orthogonal to the D3 brane are self-dual, the masses of
the $3-7$ and $7-3$ strings depend only on the distance between
the D3 and D7 branes, and  the system has $\mathcal{N}=2$ unbroken
supersymmetry. In the case when the D7 fluxes are not self-dual,
supersymmetry is spontaneously broken and the masses of the
stretched strings have an additional contribution proportional to
the amount by which the worldvolume fluxes are not self dual
($(\theta_1- \theta_2)^2$ in the notation of \cite{DHHK}). The
sign of this contribution depends on the charges of the strings:
one string gets a negative contribution and the other one gets a
positive contribution, in addition to the positive mass coming
from the distance between branes. As a result, the positively
charged string may become tachyonic at the critical point where
the distance between the branes precisely cancels the contribution
due to the non-self-dual flux on the D7 brane. This allows one  to
identify the D3/D7 brane system with the $\mathcal{N}=2$ version
\cite{K01,pterm} of the gauge theory of the D-term inflation model
\cite{Binetruy:1996xj}, with one hypermultiplet
\cite{DHHK} and a Fayet-Iliopoulos (FI) term.

The splitting of mass in the hypermultiplet leads to the logarithmic
correction to the gauge theory potential, which can also be
reproduced by the calculation of the one-loop amplitude for the
effective interaction between the branes. This field theory
one-loop correction will have an additional factor in case that
one has more than one hypermultiplet.

The end of the waterfall stage of standard D-term  hybrid
inflation \cite{L}
 solves the D-flatness
condition, and one ends up with  an  unbroken supersymmetry and a
spontaneously broken $U(1)$ gauge symmetry. In the D3/D7 brane
system, this final state corresponds to  a bound state, in which
 the  D3 brane is dissolved as a deformed Abelian instanton on the $U(1)$ DBI
theory on the D7-brane such that the Chern-Simons coupling on the
D7 brane is \footnote{For the compact case, i.e. when we have $K3
\times T^2/\mathbb{Z}_{2}$, we also need to incorporate the
contribution of the A-roof genus terms in $S_{CS}$. In that case,
the instanton number $\frac{1}{16\pi^{2}}\int_{K3}
\mathcal{F}\wedge\mathcal{F}$ should be
$(\frac{1}{16\pi^{2}}\int_{K3} {\rm tr} ~{\cal R} \wedge {\cal R}
+1)$, where ${\cal R}$ is the curvature two form.}
\be S_{CS}= {1\over 16 \pi^2} \int_{D7}C_{4}\wedge {\cal F} \wedge
{\cal F}= \int_{D3}C_{4}. \label{CS1}\ee Here, ${\cal F}_{mn}=
(dA-B)_{mn}$ has a contribution from both the D7 worldvolume gauge
field and the background  NSNS two-form   (in the directions orthogonal
to the D3). The  NSNS two-form      plays the role of a noncommutative
deformation parameter in this context. The anti-self-dual part of
the deformation parameter is related to the constant part of
$({\cal F}- \ast {\cal F})$ on the D7-brane. Only when the deformation
parameter is non-vanishing does the  Abelian instanton solution
have finite energy \cite{seibergwittenncg}. The D3/D7 bound state
has an unbroken supersymmetry, derived from $\kappa$-symmetry on
the D7-brane. The endpoint vacuum is therefore described by a
non-marginal bound state of D3 and D7-branes that corresponds to
the Higgs phase of the gauge theory of the D3/D7 system with the
FI term provided by the nonvanishing $\theta_1- \theta_2$.

The low energy effective action (i.e. to leading order in $g_s$
and $\alpha'$) for a single D3 brane and a single D7 brane in flat space is
\begin{eqnarray} \label{tdualedaction}
S&=&  \int d^8x \left[-\frac{1}{4g_7^2}F_{S T}F^{S T} -\frac{1}{2
g_7^2 (2\pi\alpha')^2}\left((\partial_{\mu} X^i)^2+(\partial_m
X^i)^2\right) \right] \nonumber
\\
&&+ \int d^4x \left[-\frac{1}{4g_3^2}F'_{\mu \nu}F'^{\mu \nu}
-\frac{1}{2 g_3^2 (2\pi\alpha')^2}\left((\partial_\mu X'^i)^2 +
(\partial_\mu Y'^m)^2 \right) \right]
\nonumber \\
&&+ \int d^4x\left[-|D_{\mu}\chi|^2-\left( \frac{X^i-X'^i}{2 \pi
\alpha'}\right)^2 |\chi|^2
-\frac{g_3^2}{2}(\chi^{\dagger}\sigma^A\chi)^2 \right] +
S_{CS}\quad .
\end{eqnarray}
The indices are as follows: $\mu$ runs from   0 through 3; $i$ denotes
directions 4 and 5; $m$ extends from  6 to 9;  and
 $S, T$ will include both the $\mu$
and $m$ directions. The primed (unprimed) fields are the light open
string degrees of freedom on the 3-brane (7-brane). The $\chi$
doublet arises as the lightest degree of freedom from stretched
strings between the D3-brane and the parallel D7-brane. That $\chi$
is generically heavy is indicated by its mass which is proportional
to $(X^i-X'^i)^2$. The covariant derivative is $D_{\mu}\chi =
(\partial_{\mu}+iA_{\mu}-iA'_{\mu})\chi $. Note that $\chi$ has
units $L^{-1}$ appropriate for a scalar in four dimensions. Note
also that the fields $X^i$ have units of $L$, and $A_{\mu}$ has
units of $L^{-1}$ in all dimensions. Recall that $g_p^2 =
(2\pi)^{p-2}g_s \alpha'^{(p-3)/2}$. The last term contains the
various Chern-Simons terms on the two branes.

We would like to place this system in a background that has some
$B_{mn}$ field, a Ramond-Ramond $F_5=dC_{(4)}$ turned on
(consistent with the underlying orbifold as well as orientifold
operation) and has
the curved metric\footnote{In this section,
we are only compactifying on the K3 directions so as to
avoid having to cancel the charges of the 7-branes and 3-branes.
Note that  we are
also ignoring the warp factors arising from the back-reactions of
branes and fluxes on the geometry. In the next section we will consider a
fully compact scenario.}
\begin{equation}\label{metric}
ds^2=R^{-6}(x)g^E_{\mu\nu}dx^{\mu}dx^{\nu} +
R^2(x)(ds^2_{\mathbb{R}^2/\mathbb{Z}_2}+ds^2_{K3}) \quad .
\end{equation}
Here, $g^E_{\mu\nu}(x)$ is the four-dimensional metric in the
Einstein  frame. The function $R(x)$ is a function of the noncompact
directions $x^{\mu}$ and is to represent the overall length scale of
the compact $K3$\footnote{Here, for simplicity, we keep track of only one of the
several K\"ahler moduli of the compact space.}. 
The powers of $R(x)$ in each part of the metric
are chosen so that the four dimensional gravitational action is in
Einstein frame \cite{Burgess:2003ic}. The function $R(x)$ should be
thought of as a very weak function of $x$ (we will drop its
derivatives) which we would aim to stabilize by wrapping branes on
some cycles in the fully compact $K3\times T^2/\mathbb{Z}_2$,  invoking the
mechanisms of \cite{KKLT}.

\subsection{The D7 action}
First, we will  consider    the D7 part of the action. In the curved background
(\ref{metric}), the first line of (\ref{tdualedaction}) modifies to
\begin{eqnarray}
S_{D7}&=& \int d^4x d^4y \sqrt{-g_E}R^{-12}\sqrt{g_{K3}}R^4
\left[-\frac{1}{4g_7^2}(F_{\mu\nu}F_{\rho\sigma}g_E^{\mu\rho}g_E^{\nu\sigma}R^{12}
+ {\cal F}_{mn}{\cal F}_{rs}g_{K3}^{mr}g_{K3}^{ns}R^{-4})
\right. \nonumber \\
& & \left. -\frac{1}{2 g_7^2 (2\pi\alpha')^2}\partial_{\mu}
X^i\partial_{\nu} X^j g_E^{\mu\nu} g^{\mathbb{R}^2}_{ij}R^8 \right]
+ \mu_7 \frac{(2 \pi \alpha')^2}{2!}\int  C_{(4)}\wedge {\cal F}
\wedge {\cal F} \ ,
\end{eqnarray}
where we have neglected the fluctuations of $X_i$ in the $K3$
directions and exhibited the powers of $R(x)$ explicitly\footnote{As mentioned
earlier, there is an additional coupling coming from the
${\rm tr}~{\cal R} \wedge {\cal R}$
term on the seven branes. This coupling comes with a relative minus sign. We will
henceforth assume that we have chosen the instanton numbers in such a way as to cancel this
contribution. As a bonus, this will also avoid the issues of enhan\c{c}ons
 etc. \cite{JPP} in the system.
For the compact case, which we discuss later, these complications will not be there.}.
The covariant 2-form on the D7 brane consists of 2
parts \be {\cal F}\equiv F-B\equiv dA- B . \ee Here, $F_{mn}$ is the
field strength of the vector field $A_m$ which lives on the brane
and $B_{mn}$ is the pullback of the space-time NSNS two-form field
to the worldvolume of the D7 brane. We have also added the
relevant Chern-Simons piece induced by the background RR field.
Now we take $\int d^4y \sqrt{g_{K3}} = V_{K3}$ to be the volume of
some fixed $K3$. Carrying out the integral over the $K3$ then
yields
\begin{eqnarray}
S_{D7}&=& \int d^4x  \sqrt{-g_E}
\left(-\frac{1}{4\tilde{g}_3^2}(F_{\mu\nu})^2 -\frac{R^{-4}}{2
\tilde{g}_3^2 (2\pi\alpha')^2}(\partial_{\mu} X^i)^2  \right)
\nonumber
\\
& & +\int Vol_{(4)} R^{-12}
\left(\frac{-1}{4g_7^2}\right)\int_{K3} {\cal F} \wedge *{\cal F}
+ \frac{1}{4\pi g_s g_7^2} \int C_{(4)} \int_{K3} {\cal F} \wedge {\cal F}
\quad,
\end{eqnarray}
where the coupling constants are
\begin{equation} \frac{V_{K3} R^4}{g_7^2} = T_7 (2 \pi \alpha')^2 V_{K3} R^4 =
\frac{V_{K3}R^4}{(2\pi)^4\alpha'^2}\frac{1}{g_3^2}=\frac{1}{\tilde{g}_3^{2}}
\quad .
 \label{couplingconstant}
\end{equation}
Note that $C_{(4)}$ is part of the background, $not$ from the D3.
The ${\cal F}$ factors in this expression are along $K3$, and
$C_{(4)}$ is extended in the flat, non-compact directions. The
background four-form is $C_{(4)}=g_s \pi R^{-12}
Vol_{(4)}/2 $. The Chern-Simons term then combines with the
kinetic term  for the gauge field in the $K3$ directions to make
the second line
\begin{equation}
\int Vol_{(4)} R^{-12}\left(\frac{-1}{8g_7^2}\right)
\int_{K3}{\cal F}^{-}\wedge *{\cal F}^{-} \quad ,
\end{equation}
where $( {\cal F}-*{\cal F})\equiv {\cal F}^-$. Here,  the $*$ is the
Hodge star on the fixed volume $K3$. One may think that $C_{(4)}$
would not point in the non-compact directions, but recalling that
the  $SL(2,\mathbb{Z})$ invariant five form needs to be self
(Hodge) dual in ten dimensions tells us that there's $C_{(4)}$ in
all ten directions. Either way, the action is now
\begin{eqnarray}
S_{D7}&=& \int d^4x  \sqrt{-g_E}
\left(-\frac{1}{4\tilde{g}_3^2}(F_{\mu\nu})^2 -\frac{R^{-4}}{2
\tilde{g}_3^2 (2\pi\alpha')^2}(\partial_{\mu} X^i)^2  \right)
\nonumber
\\
& & -\int Vol_{(4)} R^{-12}\frac{1}{8g_7^2} \int_{K3} {\cal
F}^{-}\wedge *{\cal F}^{-}\quad .
\end{eqnarray}
If one wanted to add more coincident D7s at this point, this
expression would be suitably generalized by making the connection
$A_\mu$ a $U(N_7)$ gauge field, letting the field $X^i$ be
$U(N_7)$ adjoint valued, giving it the appropriate covariant
derivative and adding an overall trace to the entire expression.

\subsection{ The D3 and stretched strings}

Now let's put the D3 on the same metric and RR background. The
expression becomes

\begin{equation}
S_{D3}  =  \int d^4x \sqrt{-g_E}\left[ \frac{-1}{4g_3^2}(F'_{\mu
\nu})^2  - \frac{R^{-4}}{2 g_3^2 (2\pi\alpha')^2}\left(
(\partial_\mu X'^i)^2 + (\partial_\mu Y'^m)^2 \right) \right]
\quad .
\end{equation}
As far as the hypermultiplet is concerned, there are now three
modifications with respect to the simple flat space
 action  (\ref{tdualedaction}): \\
(i) The mass
for the hypers (from $(X^i-X'^i)^2$) picks up some factors of $R$.\\
(ii)  The $g_3^2$ that appears in front of the D-term changes
since the hypers are also charged under the gauge field on the D7
\footnote{Note that
 this contribution is not present when the K3
is of infinite volume, as in (\ref{tdualedaction}).}.\\
 (iii) There is also an overall factor of
$R^{-12}$.\\
 Putting everything together, the new action is
\begin{eqnarray}  \label{dimreducedaction}
S & = & \int d^4x  \sqrt{-g_E}
\left[-\frac{1}{4\tilde{g}_3^2}(F_{\mu\nu})^2 -\frac{R^{-4}}{2
\tilde{g}_3^2 (2\pi\alpha')^2}(\partial_{\mu} X^i)^2 -
R^{-12}\frac{1}{8g_7^2} \int_{K3}{\cal F}^- \wedge *{\cal F}^- \right. \nonumber \\
& &  - \frac{1}{4g_3^2}(F'_{\mu \nu})^2  - \frac{R^{-4}}{2 g_3^2
(2\pi\alpha')^2}\left( (\partial_\mu X'^i)^2 + (\partial_\mu
Y'^m)^2 \right) \nonumber \\
& & \left. -R^{-6}|D_{\mu}\chi|^2-R^{-10} \left( \frac{X^i-X'^i}{2
\pi \alpha'}\right)^2 |\chi|^2 -\frac{R^{-12}(g_3^2 +
\tilde{g}_3^2)}{2}(\chi^{\dagger}\sigma^A\chi)^2 \right] \ .
\end{eqnarray}
In this equation, the hypermultiplet covariant derivative is still
$D_{\mu}\chi = (\partial_{\mu}+iA_{\mu}-iA'_{\mu})\chi $. We
perform  a gauge field redefinition
\begin{equation}
gW_\mu=A_\mu - A'_\mu \qquad gW'_\mu=\frac{g_3}{\tilde{g}_3}A_\mu
+\frac{\tilde{g}_3}{g_3} A'_\mu \qquad g^2=g_3^2+\tilde{g}_3^2
\end{equation}
under which $\frac{1}{\tilde{g}_3^2}F^2 + \frac{1}{g_3^2}F'^2 =
F_W^2 + F_W'^2$. The scalars must also be rotated,
\begin{eqnarray}
& g S=\frac{R^{-2}}{2\pi\alpha'g\sqrt{2}}\left((X^4-X'^4)+i(X^5-X'^5)\right) & \nonumber \\
& gS' =
\frac{R^{-2}}{2\pi\alpha'g\sqrt{2}}\left((\frac{g_3}{\tilde{g}_3}X^4+\frac{\tilde{g}_3}{g_3}X'^4)
+ i(\frac{g_3}{\tilde{g}_3}X^5+\frac{\tilde{g}_3}{g_3}X'^5)
\right) &
\end{eqnarray}
after which their kinetic term has the standard form. In order to also
bring $\chi$ into the standard form,
we must rescale, $\chi \to R^3\chi$. The action
at this point is
\begin{eqnarray}  \label{canonicallynormed}
S & = & \int d^4x  \sqrt{-g_E} \left[-\frac{1}{4}(F_W)^2
-\frac{1}{4}(F_{W'})^2 -|\partial S|^2 -|\partial S'|^2-
R^{-12}\frac{1}{8g_7^2} \int_{K3}{\cal F}^- \wedge*{\cal F}^- \right. \nonumber \\
& & \left. -|D_{\mu}\chi|^2- 2g^2|S|^2 |\chi|^2 -\frac{(g_3^2 +
\tilde{g}_3^2)}{2}(\chi^{\dagger}\sigma^A\chi)^2 \right] \ ,
\end{eqnarray}
where we have ignored the motion of the $D3$ brane
parallel to the wrapped direction of the $D7$ brane. We have
also neglected derivatives of $R(x)$. With these gauge fields,
$D_{\mu}\chi = (\partial_{\mu}+igW_\mu)\chi $, and all fields have the standard normalization
of ${\cal N}=1$ chiral superfields.

\subsection{ P-term action}

Now since this system is known to have rigid $\mathcal{N} =2$ supersymmetry
(broken to $\mathcal{N}=1$ when coupled to gravity)
in $d=4$, there must be some way to express this action in terms
of standard $\mathcal{N} =1$ language. The suggested expression in
terms of (rigid) $\mathcal{N}=1$ chiral superfields is the so
called P-term model \cite{pterm}
\begin{eqnarray} \label{susyL}
\mathcal L & = & -\frac{1}{4}F_{\mu \nu}^2 -|D_{\mu}\Phi_+|^2 -|D_{\mu}\Phi_-|^2 - 2 g^2
\left(|S|^2 (|\Phi_+|^2 + |\Phi_-|^2) + \Bigl|\Phi_+ \Phi_- -
\frac{\xi_+}{2}\Bigr|^2\right) \nonumber
\\
& & -\frac{g^2}{2}(|\Phi_+|^2 - |\Phi_-|^2 - \xi_3)^2
\quad .
\end{eqnarray}
In this expression, the fields are as follows. The complex fields
$\Phi_{+}$ and $\Phi_{-}$ sit in two chiral multiplets of opposite charge with
covariant derivatives
 $D_{\mu}\Phi_{\pm}=(\partial_{\mu} \pm ig
W_{\mu})\Phi_{\pm}$. The field $S$ is neutral under the gauged
$U(1)$ and likewise sits in an $\mathcal{N}=1$ chiral multiplet.
In  $\mathcal{N}=2$   language,
the pair $(\Phi_+,\Phi_-)$ should be viewed as an
$\mathcal N$=2   hypermultiplet charged under a $U(1)$ gauged by
the    $\mathcal{N}=2$
vector multiplet  $(W_\mu,S)$. $\Phi_+$ and $\Phi_-$
form a doublet under the  $\mathcal{N}=2$    R-symmetry group
   $SU(2)_R$.
In   $\mathcal{N}=1$ language, the     potential can be obtained from the
superpotential $W=\sqrt{2}gS(\Phi_+ \Phi_- -\xi_+/2)$ and D-term
$D= |\Phi_+|^2-|\Phi_-|^2 - \xi_3$. The coupling constant $g$ in
the superpotential is the same as the $U(1)$ gauge coupling so
that the system is in its $\mathcal N$=2 limit. The $U(1)$
Fayet-Iliopoulos term $\xi_3$ has been included. Note that the
fields in the P-term expression are chiral superfields with the standard kinetic terms of the type
$D_{\mu}\Phi D^{\mu}\Phi^*$.
\subsection{Comparing the actions}

The action in eq. (\ref{canonicallynormed}) is to be compared with
the P-term action, eq. (\ref{susyL}). Since the fields in the brane expression
are already normalized as  $\mathcal{N}=1$ chiral fields,
the comparison is straightforward. The neutral
scalar $S$ is to be identified in the two expressions. The hypers
should be identified as $\chi_1=\Phi_+$ and $\chi_2=\Phi_-^*$ so
that the covariant derivative behaves correctly,
\begin{equation}
|D_{\mu}\chi|^2 = |(\partial_{\mu}+igW_\mu)\chi|^2 \to
|(\partial_{\mu}+igW_{\mu})\Phi_+|^2 +
|(\partial_{\mu}-igW_{\mu})\Phi_-|^2 \ .
\end{equation}
Just as in the P-term action in section 2.3,
the fields $(W_\mu,S)$ should be thought of as an
$\mathcal N$=2 $U(1)$ vector multiplet. The other gauge multiplet,
$(W'_\mu,S')$ has completely decoupled and is irrelevant in the
P-term model. From the brane picture, it is the decoupled center
of mass degree of freedom for the total D3-D7 system.

If we concentrate on the $\xi_\pm=0, \xi_3=\xi$ case of the P-term
model, we know that during the inflationary phase (Coulomb phase),
supersymmetry should be broken by the FI term in the D-term. On
the brane side, we know from kappa symmetry that non-selfdual
fluxes on the D7 gauge field in the $K3$ directions break
supersymmetry with the same mass splittings \cite{DHHK}. Thus one finds
that the $(\mathcal{F}-*\mathcal{F})^2$ term in eq.
(\ref{canonicallynormed}) is actually the $g^2\xi^2/2$ term of the
P-term model. Namely,
\begin{equation}
R^{-12}\frac{1}{8g_7^2} \int_{K3}{\cal F}^- \wedge *{\cal F}^- \to
\frac{g^2\xi^2}{2} \quad . \label{FFI}
\end{equation}
The $R^{-12}$ dependence is the same as was found in
\cite{Burgess:2003ic}. Note that the last term of
(\ref{canonicallynormed}), when expanded, actually contains some
of the terms arising from the superpotential in the P-term
Lagrangian.

The above analysis more or less summarizes the precise dictionary
between the brane action and the $P$-term action of \cite{pterm}.
However, as we discussed earlier, the analysis is only for a
compact $K3$ and as such would require a more detailed structure
when we try to compactify the $\mathbb{R}^2/\mathbb{Z}_2$ to a
$T^2/\mathbb{Z}_2$. In the next section we will carry out the
compactification of the directions orthogonal to the 7-branes in
the full F-theory picture which should also capture all of the
non-perturbative corrections to the system. Furthermore we will also
generalize the above configuration to incorporate
enhanced global and local symmetries by
considering stacks of $D7$ and $D3$ branes on top of each other.

\section{Generalized $D3/D7$ model with additional global or local symmetries}

In the previous
 section, we saw how the Lagrangian for a system of a single $D3$ and a single
D7 can be derived using the world volume dynamics of the individual branes. We also discussed
the additional contributions to the action when hypermultiplets are added to the system. The
Lagrangian that we get is precisely the expected one, studied earlier in
\cite{pterm}.

All the discussion  we had  in Section 2,
  however, was
 only for non-compact $\mathbb{R}^2/\mathbb{Z}_2$, i.e the orthogonal
directions to the $D7$ (and also the $D3$) branes were still taken
to be  non-compact (the $K3$ directions, however,  were already
assumed compact in Section 2).
As we know, the non-compact cases are
only an approximation to the compact case
studied earlier in \cite{DHHK}.
Going to the compact case will mean
 to consider the orthogonal space as a $\mathbb{P}^1$
or $T^2/{\cal I}_2$ (where ${\cal I}_2$ is the orbifold
action that reverses the two direction of the $T^2$ torus)
on which both the $D7$ and the $D3$ branes are
points\footnote{Recall that the compact $\mathbb{P}^1$ appears from the F-theory picture that we discussed in \cite{DHHK}. In terms of the
F-theory four-fold, this is simply a $K3 \times K3$ compactification with fluxes.}.
Of course, charge conservation requires that we should have no global seven brane charges
 floating around, which,
in terms of the underlying F-theory construction requires us to
view the system as a Weierstrass equation \cite{vafaF} \be
\label{weier} y^2 = x^3 + x F(z) + G(z) \ee with $z$ as the
coordinate of the $\mathbb{P}^1$. The axion-dilaton of type IIB
forms the modular parameter $\tau$ of the torus that is determined
by the degree eight and degree twelve polynomials $F$ and $G$
respectively using the $J$ function. All these have been defined
earlier in \cite{DHHK} so we will not discuss them further here.
The global monodromy (which determines the global seven brane
charges) is zero and the $D3$ brane charge is cancelled by
switching on NSNS and RR three-forms satisfying the equation \be
\label{chargec} 2i~\int_{\cal M} {\rm Im}~\tau~\omega_1 \wedge
\omega_2 = 23 \ , \ee where we have chosen a single probe $D3$
brane on the manifold ${\cal M} = K3 \times T^2/{\cal I}_2$. The
relation between $\omega_1$ and $\omega_2$ to the three forms
$H_{NS}$ and $H_{RR}$ is given in \cite{DHHK}. {}From  the F-theory
point of view,    the underlying four-fold has an Euler number of
576. We are also assuming that the $D3$ brane is located at a
point ${\tilde z}_1$ on the $\mathbb{P}^1$, whose coordinate is
$z$.

In the above relation (\ref{chargec}), if $\omega_1$ and
$\omega_2$ are arbitrary, then in general supersymmetry is broken.
This would imply that the $D3$ brane is moving towards the $D7$
branes. In \cite{DHHK} we called this phase  the {\it Coulomb}
phase of the hybrid inflation. In the situation where we have
isolated one $D7$ from the bunch of  seven-branes,    the $D3$
will eventually fall into the $D7$ brane as an instanton. Because
of the presence of FI terms, i.e, because of the gauge field
${\cal F}$, this instanton will actually be a non-commutative
instanton. The condition for this final state to preserve
supersymmetry can be described as follows. If one isolates from
the F-theory curve (\ref{weier}) one of the singularities  of $F$
and $G$, at which the torus  fiber (parameterized by a complex
coordinate $\zeta$) degenerates {\it once},
 and  if one assumes
  the existence of a single normalizable (1,1) form $\Omega$
in the neighborhood of this singularity,
then supersymmetry will eventually be restored if the
four-form
\be
\label{foform}
 {\cal F} \wedge \Omega - \omega_1 \wedge d\zeta - \omega_2 \wedge d{\bar\zeta} \ee
is self-dual. Note that the self-duality of the four-form
(\ref{foform}) now does not guarantee the self-duality of the
seven brane gauge fluxes, ${\cal F}$, i.e, we have in general
${\cal F} \ne *{\cal F}$, where the Hodge star is with respect to
 the underlying four-dimensional $K3$
base\footnote{The fact that there exists
a solution   with self-dual four form   (\ref{foform})
  is highly non-generic and has
  been discussed earlier in \cite{drs}.}.
We have also
identified the one forms
\be
\label{omeforms}
(dX, dY) \equiv \left({\tau d{\bar\zeta} - {\bar\tau} d\zeta \over 2i {\rm Im}~\tau}, ~~
{d\zeta - d{\bar\zeta} \over 2i {\rm Im}~\tau}\right)
 \ee
as determining the $T^2$ fiber of our F theory construction. For
more details, the reader may want to consult \cite{vafaF}. In our
earlier paper \cite{DHHK}, this final stage, when the probe $D3$
brane falls into the $D7$ brane as a non-commutative instanton,
was called  the {\it Higgs} phase of the hybrid inflation. To
compare this situation to the case {\it without} any incoming $D3$
brane, observe that in that situation supersymmetry would be
restored when $\omega_2 = \ast \omega_1$ (the Hodge star now being
with respect to the six dimensional compact manifold). Of course,
for our case, since we will always have the probe $D3$ brane in
the picture, this situation will never arise. Further details on
this has appeared in \cite{drs}.

For the compact case that we discussed earlier in \cite{DHHK} the
metric for the system can be easily written down taking into
account   {\it all} the effects of the backreactions of the branes
and fluxes on geometry. Following \cite{drs}, we write the metric
for the system as:
\be \label{metiib} ds^2 = \Delta^{-1}
ds^2_{0123} + \Delta (ds^2_{\mathbb{P}^1} + ds^2_{K3}) \ , \ee
where
$\Delta$ is the warp factor. The metric looks similar to the
metric for a single $D3$ brane at a point on the compact manifold
${\cal M}$, because the fluxes and branes simulate the situation
of having $D3$ branes at a point on ${\cal M}$. The solution to
the warp factor is a little involved, because we now require the
situation where we have isolated (a) one $D7$ from the whole bunch
of seven branes, or
 (see   the next subsection)
 (b) two $D7$ branes from the bunch of seven branes.
The second case (b) is actually what we are mostly interested in
this paper, because, as discussed in earlier sections, this will
give rise to two hypermultiplets on the $D3$ brane and also global
symmetries in the system. Before we go into the issue of the  warp
factor, let us therefore see how global symmetries arise in the
compact model from F-theory.

\subsection{Global symmetry from two $D7$ branes}

There are two issues that arise when we want to incorporate global
symmetries to the effective world volume theory on the $D3$ brane.
As we will generate these  global symmetries from the gauge
symmetries on coincident  $D7$ branes, the first issue is the
decoupling of $D7$ brane gauge dynamics from the $D3$ brane
theory. However, what would suffice for our case will be to make
the $D7$ gauge coupling constant sufficiently small compared to
the $D3$ brane gauge coupling. This is possible for our case
because the effective $D7$ coupling is suppressed by the volume of
the underlying compact $K3$ manifold  (see
eq.(\ref{couplingconstant})). Therefore when we determine the
covariant derivatives on the $D3$ brane world volume, we can
effectively neglect the interactions due to the $D7$ gauge fields, because
the $K3$ manifold is assumed to be fixed at a large volume.
On the other hand, the kinetic term of the $D7$ brane will
survive, and, bearing in mind the correspondence (\ref{FFI}),
 we can use this to study the  FI
  and other relevant terms for our system.

Secondly, as the existence of the global symmetry is related to the
existence of a {\it gauge} symmetry on the $D7$ branes, this
should appear from the Weierstrass equation that we discussed in
(\ref{weier}). Let us see how this works  in more detail. First,
observe   that this situation is {\it not} the constant coupling
scenario of \cite{senF, dm}. What we require is that the
discriminant $D$ have a double zero at a point $z_{1}\in
\mathbb{P}^1$: \be \label{discr} D \equiv 4F^3 + 27 G^2 = (z -
z_1)^2 H(z) = 0 \ ,
 \ee
where $H(z)$ is a generic polynomial of degree 22, which is
regular at $z_{1}$. This would imply that, for a contour $C(z_1)$
surrounding the point $z_1$, we will have: \be \label{cont}
\oint_{C(z_1)} d\tau = 2. \ee The requirement on the discriminant
$D$ in (eq. \ref{discr}) leaves the freedom to choose the
functions $F$ and $G$. Although there are possibilities of various
choices that allow a double zero, one particular set that will be
consistent for our scenario would be: \be \label{fang} F(z) =
(z-z_1)~{\tilde F}(z), ~~~~~~~~ G(z) = (z - z_1)~{\tilde G(z)},
\ee where ${\tilde F}$ and ${\tilde G}$ are polynomials of degree
seven and eleven respectively. This means that the F-theory curve
near the point $z = z_1$ will be: \be \label{aone} y^2 = x^3 +
\alpha x {\tilde z} + \beta {\tilde z}, \ee where ${\tilde z}$ is
now ${\tilde z} = z - z_1$ and we have absorbed ${\tilde F}$ and
${\tilde G}$ in the definition of $\alpha$ and $\beta$
respectively. {}From the above relations and from the
parametrizations used in \cite{tate},
 we see a local $A_1$ singularity.\footnote{This curve was
also observed in \cite{witten} in ${\cal N} = 2$
gauge theory with two flavors.
This is basically related
to the Argyres-Douglas point \cite{ad}.} This local singularity
will then appear on the $D3$ brane as a global $SU(2)$ symmetry.
The fact that there could be an $SU(2)$ symmetry that doesn't lie
on the constant coupling moduli space of F-theory was also
observed in \cite{dm}.   In fact,    the coupling $\tau$ at the
point $z = z_1$ can be seen to be \be \label{copco} \tau \simeq {2
\over 2\pi i} {\rm ln} (z - z_1). \ee
Observe now that as $z \to
z_1$, $\tau \to i \infty$ which corresponds to weak coupling, and
therefore quantum corrections will {\it not} modify this behavior.
Thus this configuration will survive perturbative as well as
non-perturbative corrections.

There are other interesting issues here. As we observed above, the
global symmetry is governed by the F-theory singularities.
Therefore, if we choose the singularities such that the symmetry
is now $E_n$ then we can have theories that have exceptional
global symmetries, perhaps giving rise to exceptional semi-local
strings. For example, a simple way to get $E_8$ as a global symmetry is
to allow the following behavior of the discriminant $D$: \be
\label{excepD} D = (z - z_1)^{10} H(z) \ ,
 \ee
where $H(z)$ is now a polynomial of degree 14. As discussed in
\cite{dm}, this can be realized at a constant coupling moduli
space of F-theory by keeping ten seven branes at one point. The
coupling near the system is strong and therefore not realized as a
simple orientifold model. We will however not go along those
interesting directions in this paper. The possibility of the
existence of semi-local strings in theories with exceptional
global symmetries
 were hinted at  earlier in \cite{HHKV},
although no concrete realization of this has yet appeared in the
literature. For the $D3/D7$ system, we see that such a scenario may be
possible to realize, as most of the global  symmetries    come from the
Weierstrass equation in F-theory \cite{dm}.
More details will appear
elsewhere.

Before moving ahead, let us summarize the content of this section
in terms of an effective gauge theory: In this section we have
described a D-brane picture with a local $U(1)$ gauge symmetry and
a global $SU(2)$ symmetry. This should now be compared with the
discussion that we had in section 2 for a $U(1)$ gauge theory and
a single hypermultiplet. In terms of an effective  4-dimensional
gauge model, we will have the following potential: \bea V&=&2
g^2\left|\Phi_+\Phi_--\tilde\Phi_-\tilde\Phi_+\right|^2+
\frac{g^2}{2}\left[|\Phi_+|^2+|\tilde\Phi_+|^2-|\Phi_-|^2-|\tilde\Phi_-|^2-\xi
\right]^2+ \nonumber\\ & &2 g^2
|S|^2\left[|\Phi_+|^2+|\Phi_-|^2+|\tilde\Phi_+|^2+|\tilde\Phi_-|^2\right].
\label{potPhi} \eea where ($\tilde\Phi_+, \tilde\Phi_-$) form the second hypermultiplet.
We are also making a choice $\xi_+=0$ and
$\xi_3=\xi\neq 0$,  with regard to eq. (\ref{susyL}). This
corresponds to a choice of the canonical basis for the 2-form
${\cal F}$ when only ${\cal F}_{67}$ and ${\cal F}_{89}$ are
non-vanishing, so that $({\cal F}_{mn}^-)^2$ is given by $({\cal
F}_{67}- {\cal F}_{89})^2$. In terms of the brane language used in
section 2, the above potential can be easily derived from the $D7$
brane and the $D3$ brane actions simply by taking traces over the
adjoint representations of the gauge fields and by neglecting the
$SU(2)$ D-terms.

\subsection{Enhancement of local symmetries}

So far, we have been taking one $D3$ brane probing a system of
$D7$ branes. There is another interesting scenario where we put
two $D3$ branes probing the region near a single $D7$ brane in the
compact model. Of course, since the coupling of the $D3$ brane
cannot be ignored compared to the coupling of the $D7$ brane, we
cannot regard this as enhancing the global symmetry. The global
symmetry remains $U(1)$, but  the gauge symmetry will now
increase. For a single $D3$ probe, when it is very near the $D7$
brane and $O7$ plane, we know that the gauge symmetry is $SU(2)$
broken to $U(1)$ at all points on the moduli
space\footnote{Classically the gauge symmetry is $SU(2)$ at the
point where Higgs vev is zero. Quantum mechanically there is never
an $SU(2)$ enhancement \cite{sw}.}. This $SU(2)$ symmetry is
actually $USp(2)$ and therefore doubling the number of $D3$ branes
will make the symmetry $USp(4)$,\footnote{Again, far away from the
orientifold plane and $D7$ brane the gauge symmetry is $SU(2)$,
similar to the $U(1)$ case for the single $D3$ brane.} and, more
generally, $USp(2k)$ for $k$ $D3$ branes. Thus, classically the
symmetry is $USp(4) \times U(1)$. The former being a local gauge
symmetry and the latter global. The dynamics of this system under
full perturbative and non-perturbative corrections has been
discussed in \cite{aharony}.

To see the hypermultiplets in this setup we can use fundamental
strings stretching between the branes. In Figure 1, we see that
there are two different possibilities. In one of the
possibilities, depicted as a string stretched directly between two
$D3$ branes and the $D7$ brane
\begin{figure}
\centering
\epsfig{figure=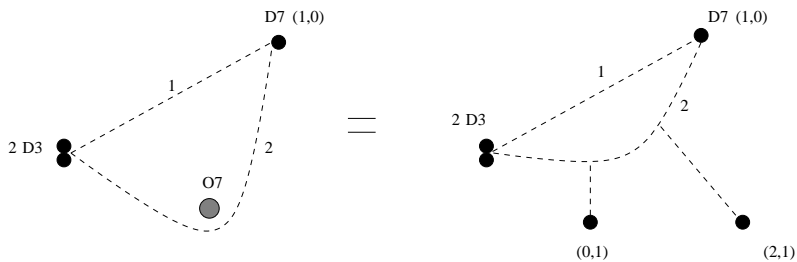} 
\begin{center}
\caption{The hypermultiplets from two $D3$ branes and one $D7$ brane.}
\label{fig-B0}
\end{center}
\end{figure}
\noindent one obtains two hypermultiplets as desired. For
notational purposes we will call this  state 1. But there are
other possibilities for the compact case where, due to global
seven brane charge conservation, we require definite numbers of
orientifold planes and seven branes. Now we can have a string that
goes all the way across an orientifold plane and comes back to the
$D7$ brane (we call this state 2 in the figure). Under
 non-perturbative corrections,   the $O7$ planes will in general split
into pairs of seven branes which we denote as (0,1) and (2,1)
seven branes, the numbers being the charges measured with respect
to axion and dilaton, respectively (in this notation, a $D7$ brane
will be denoted as (1,0)). The splitting of the orientifold plane
into seven branes create branch cuts on the $\mathbb{P}^1$.
Therefore the strings reaching the $D7$ brane undergo changes in
their charge distribution \cite{GZ}. If we call the strings
leaving from $D3$ as $\pmatrix{1 \cr 0}$, then the condition for
them to reach the $D7$ brane will be \be \label{reach}
M_{[2,1]}\cdot M_{[0,1]}\cdot \pmatrix{1 \cr 0} = \pmatrix{-1\cr
~0}, \ee where $M_{[p,q]}$ are the monodromy matrices whose
explicit form can be taken from \cite{GZ}. One can check that this
condition is met for our case. The masses of the strings stretched
between the $D3$ branes and $D7$ brane can be calculated and will
in general be much heavier than state 1. Some of these states may
even be non-BPS. Since all these states are much heavier than the
state 1 we can effectively ignore those states for our purposes
and concentrate only on the state 1.

Before we end this section, let us mention a few words on the
cosmology of this model. One first notes that the $SU(2)$ coupling
of the two $D3$ branes (at far distance from $D7$) can no longer
be much smaller than the $U(1)$ gauge coupling of the $D7$ brane
unless the compactification volume is very small.  Hence,  the $SU(2)$
gauge symmetry can no longer be treated as effectively global as
we mentioned earlier. D-term inflation for such a more complicated
gauge group has not yet been  studied. Therefore, in this paper,
we will focus on a $D3/D7$ brane model with
two $D7$ branes and two hypermultiplets in the effective $U(1)$ gauge theory
with global $SU(2)$ symmetry.

\subsection{Analysis of the warp factor}

Now that we have seen how global symmetries and hypermultiplets
appear in our setup, it is time to go back to the issue of warp
factors that we briefly mentioned at the beginning of this
section. Since we have isolated two $D7$ branes from a system of
branes and orientifold planes, we have to carefully consider the
equation of the warp factor. In the notations that we used above,
we keep a probe $D3$ at a point $z = {\tilde z}_1$ on the
$\mathbb{P}^1$. The two $D7$ branes are located at $z = z_1$. The
metric ansatz that we used in (\ref{metiib}) involves two things:
(a) the metric on the sphere $\mathbb{P}^1$, and (b) the warp
factor $\Delta$ (we will not be too concerned with the metric of
$K3$ here). Let us first tackle the metric on the $\mathbb{P}^1$.
In terms of the  eta-function $\eta$ we can express the metric
near two $D7$ branes on top of each other as \cite{gsvy}: \be
\label{metdsev} ds^2_{\mathbb{P}^1} = g_{z \bar z} ~\vert
dz\vert^2  \simeq  {\rm Im}~ \tau(z) ~\big\vert \eta^2(\tau)~
(z-z_1)^{-{1\over 6}} dz \big\vert^2
 \ee
Putting a probe $D3$ brane and additional fluxes only introduces a
warp factor $\Delta$ to the system as discussed above in
(\ref{metiib}). The equation for the warp factor can be
approximated near the $D7$ branes as: \be \label{warpfac}
\begin{array}{c}
\displaystyle{\ast \square ~\Delta^2 = ~~ H_{NS} \wedge H_{RR} +
(4\pi^2 \alpha')^2~ \delta^2(z - {\tilde z}_1) ~ \delta^4_{K3} +} \\\\
\displaystyle{~~~~~ + {(\alpha' \pi)^2 \over 64} [{\rm tr}~({\cal
R} \wedge {\cal R}) - {\rm tr}~({\cal F} \wedge {\cal
F})]~\delta^2(z - z_1)}
\end{array}
\ee where the factor of ${1\over 64}$ comes from the gravitational
couplings on the two $D7$ branes. In writing the above relation we
have ignored the effect of other seven branes that are placed far
away from the two $D7$ branes. The position of the $D3$ brane on
the $K3$ is denoted symbolically by the delta function
$\delta^4_{K3}$.

The solution to the above equation involve various things: (a) the
choice of the three forms $H_{NS}$ and $H_{RR}$, (b) the choice of
the world volume gauge fluxes ${\cal F}$, and (c) the curvature
tensor ${\cal R}$. The full solution to this equation has not been
worked out, and therefore we will not try to address it here in
details as this would lie beyond the scope of this paper. We only
note that an approximate solution to the warp factor can be given
in terms of Green's function {\it \`{a} la} \cite{shiu}. Using
this one can show that the warp factor is typically of order 1
here as the additional factors are suppressed by the volume of the
compact manifold. Since we take large internal volume (which
earlier helped us to decouple the seven brane gauge interactions)
we can safely ignore the warping effects here. Therefore instead
of going to a more detailed analysis of the warp factor, we now
turn to the analysis of semilocal strings in this model.

\section{Semilocal BPS-strings  in  supergravity}
\label{ss:BPS}
Semilocal strings are a special class of vortex solutions that
were first found in the
extended Abelian Higgs model \cite{VA1,Hindmarsh1}.
In that  model, the Higgs sector of the conventional   Abelian Higgs model
(which gives rise to the standard  ANO-strings)
is doubled so that the theory has a global $SU(2)_g$ symmetry, in addition to
its
 local $U(1)_l$ gauge symmetry.\footnote{Simple generalizations
to $N$ Higgs fields
with a global $SU(N)_{g}$ symmetry appeared in \cite{Hindmarsh2}, whereas
\cite{HHKV} discussed other symmetry groups
that could possibly also  give rise  to semilocal defects.}
In the last few sections we saw in detail how the $D3/D7$ system realizes
easily the required global and local symmetries of this model, which
 is the main requirement for the presence    of the  semilocal strings.
However, our model has much more structure, as it is    also
embedded into $\mathcal{N}=1$ supergravity. In this section, we
make use of this extra structure and show how the semilocal
strings   can arise as $1/2$-BPS solitons  in the extended Abelian
Higgs model coupled to four-dimensional, $\mathcal{N}=1$
supergravity. While it is known that  semilocal strings can
satisfy BPS-type energy bounds in the case of critical coupling
\cite{VA1,Hindmarsh1,GORS}, the precise connection with
local supersymmetry has so far not been   established in a clear
an precise way.\footnote{For related work in three dimensions, see
\cite{Edelsteinsemilocal}.} Our treatment extends the discussion
of ref. \cite{Dvali:2003zh} for the locally supersymmetric ANO-vortex to
the semilocal cosmic string. In the analysis below, we will ignore
the effectively decoupled D7 brane gauge fields completely,
dropping their kinetic term.

\subsection{The Lagrangian}
In section 2, we derived the essential features of the low energy
effective action of a simple $D3/D7$ brane system in a $K3\times
T^{2}/\mathbb{Z}_{2}$ compactification of type  IIB string theory
(see eq. (\ref{susyL}) for  the $\mathcal{N}=2$    supersymmetric
non-gravitational part). As sketched   in section 3, this system
can be generalized to describe one $D3$ brane near two coincident
$D7$ branes with gauge group  $U(1)_{l}$ and an effectively global
symmetry $SU(2)_{g}$. This global symmetry acts by rotating two
(rigid) $\mathcal{N}=2$ hypermultiplets, $ \left( \Phi_{+},
\Phi_{-} \right) $ and $ ( \tilde{\Phi}_{+}, \tilde{\Phi}_{-} )$,
into each other\footnote{$SU(2)_g$ should not be confused with the
R-symmetry group $SU(2)_{R}$, which rotates $\Phi_{+}$ into
$\Phi_{-}$ and $\tilde{\Phi}_{+}$ into $\tilde{\Phi}_{-}$.}. The
two hypermultiplets are accompanied by an $\mathcal{N}=2$ vector
multiplet $(W_{\mu},S)$ corresponding to the gauged $U(1)_{l}$
symmetry (we are suppressing the fermions for the moment). It can
be shown that the neutral and negatively charged scalar fields
have to vanish on the string solution \cite{PRTT96,ADPU02}. So we will from now on set
in this section
\begin{equation}
S=\Phi_{-}=\tilde{\Phi}_{-}=0
\end{equation}
in equation (\ref{potPhi}).
This collapses the superpotential,
which we will henceforth set to zero.
For the two remaining, positively charged,
 scalars $\Phi_{+}$ and $\tilde{\Phi}_{+}$, we employ the somewhat
more convenient notation
\begin{equation}
\varphi_{1}= \Phi_{+}, \qquad \varphi_{2} = \tilde{\Phi}_{+} \ .
\end{equation}

Coupling this model to gravity reduces
 supersymmetry to $\mathcal{N}=1$,
and the total
field content including the fermions now  derives from
 the $\mathcal{N}=1$  supergravity
multiplet $(e_{\mu}^{a},\psi_{\mu})$, two chiral multiplets
$(\chi_i,\varphi_{i})$ $(i=1,2)$ and
a vector multiplet $(W_{\mu},\lambda)$. Clearly, the
 two chiral multiplets carry
an electric
charge $+1$ under the local $U(1)_l$ symmetry gauged by $W_\mu$
and transform as a doublet
under the global $SU(2)_g$ symmetry. In the following, we employ
the supersymmetry conventions
of ref.  \cite{BDKV,KKLV}. In
particular, we use $\varphi^{i}$ to denote the complex conjugate of
$\varphi_{i}$  and indicate the chirality
of their superpartners by
 $\chi_{i}$  (left handed)    and $\chi^{i}$   (right handed).

Taking the K\"{a}hler potential and the gauge kinetic function to
be minimal, the bosonic part of the Lagrangian then reads

\begin{eqnarray}
e^{-1}{\cal L}_{\rm bos}=-\ft12M_P^2 R -D_\mu
\varphi_i\, D^\mu \varphi^i
 -\ft{1}{4} F_{\mu \nu } F^{\mu \nu }- V^D\,,
 \label{bosonic2}
\end{eqnarray}
where the D-term  potential is defined by
\begin{equation}
V^D= \frac{1}{2} D^2 \qquad  D\, = g(\xi -  \varphi^i\, \varphi_i )  \,
\label{Pphi}
\end{equation}
and
\begin{equation}
F_{\mu\nu}\equiv {\partial }_\mu W_\nu- {\partial }_\nu W_\mu\, ,\qquad
D_\mu \varphi_i\equiv ({\partial }_\mu -\rmi g W_\mu) \varphi_i\,
\end{equation}
denote the usual Abelian field strength and the $U(1)_l$ covariant
derivative. The transformation rules of the fermions in a purely
bosonic background are
 \begin{eqnarray}
\delta \psi _{\mu L}  & = & \left( \partial _\mu  +\ft14 \omega _\mu
{}^{ab}(e)\gamma _{ab}
+\ft 12\rmi A_\mu^B \right)\epsilon_L \,, \nonumber\\
\delta \chi_i & = & \ft12(\not\! {\partial }-\rmi g \not\! W) \varphi_i
\epsilon _R\,,
 \nonumber\\
\delta \lambda  &=&\ft14\gamma ^{\mu \nu } F_{\mu \nu }\epsilon
+\ft12\rmi \gamma _5  D
 \epsilon  \,. \label{susyf}
\end{eqnarray}
Here the (composite)
gravitino $U(1)$ connection,  $A_\mu ^B$,  is given by
 \begin{eqnarray}
  A_\mu ^B&=&\frac{1}{2M_P^2}\rmi\left[ \varphi_i\partial _\mu
\varphi^i -\varphi^i \partial _\mu \varphi_i\right]
  +\frac{1}{M_P^2}W_\mu  D\nonumber\\
  &=&\frac{1}{2M_P^2}\rmi\left[ \varphi_iD_\mu \varphi^i
-\varphi^i D_\mu \varphi_i\right]
  +\frac{g}{M_P^2}W_\mu  \xi  \,.
  \label{AmuBinphi}
\end{eqnarray}

\subsection{Semilocal vortices with unbroken supersymmetry}

Our goal here is to show how the above action can give rise to
string-like  defect solution that preserve one half
of the original $\N=1$ supersymmetry. For simplicity, we restrict
ourselves to non-singular configurations that are static and
cylindrically symmetric, that is, we assume $(1+1)$-dimensional
Poincar\'{e} symmetry on the `world sheet' of the string as  well
as   a rotational symmetry with orbits orthogonal to that world
sheet. Using $(t,z)$ as the world sheet coordinates, we align the
rotation axis of the string with the $z$-direction. The orthogonal
distance from the rotation axis is denoted by $r$ and rotations
about the axis are parameterized by  an angular coordinate
$\theta$. The space-time metric can be brought to the form

\begin{equation}
  \rmd s^2= -\rmd t^2 +\rmd z^2+\rmd r^2 + C^2(r) \rmd \theta ^2\,.
 \label{tentativemetric}
\end{equation}
We use the  vierbein   with $e^1=\rmd r$ and $e^2=C( r)\rmd\theta $, which
leads to the spin connection
\begin{equation}
  \omega_r{}^{12}=0\,,\qquad
  \omega_\theta {}^{12}= -C'(r)\,.
 \label{12components}
\end{equation}

In order for the string to have finite  energy per unit length,
the potential and kinetic energy densities have to vanish sufficiently fast
at large distances $r$ from the string axis. As for the potential energy,
this means that
the scalar fields $\varphi_i$ have to approach the vacuum manifold $S^3
=\{\varphi_i:  |\varphi_1|^2+|\varphi_2|^2=\xi\}$
of the potential   $V^D$ as $r\rightarrow \infty$. A
  cosmic string in this model thus defines a mapping of the spatial   circle
$S^1_{\infty}$ at infinite $r$ into the vacuum manifold $S^3$.

If this was the only condition,
one would conclude that there could be no topologically non-trivial
 vortices, as
$\pi_1(S^3)$=\{1\}, i.e., any non-trivial mapping of $S^{1}_{\infty}$
could be deformed into the trivial map (i.e., the vacuum) at
no cost of potential energy.

As was pointed out in \cite{VA1, Hindmarsh1},    however,  the  requirement
   of finite \emph{kinetic}
energy per unit length is of equal importance, as it
implies   that $
D_\mu \varphi_i $ also vanish sufficiently fast at large $r$.
This  further
restricts $(\varphi_1,\varphi_2)$ to approach a $U(1)_l$ gauge orbit
(i.e.,  a great circle $S^1_l$) on $S^3$ as $r$ goes to infinity.
The Higgs field  $(\varphi_{1},\varphi_{2})$    at
$r \rightarrow    \infty$ thus really  defines a
map of $S^1_{\infty}$ into a gauge orbit $S^1_{l}$
 of $U(1)_{l}$ on the vacuum manifold
$S^3$. Any such map is characterized by a winding number  $n\in
\pi_{1}(S^1_{l})=\mathbb{Z}$ and an associated gauge connection that
compensates the gradient energy from the angular dependence of the
Higgs field.   Stokes's   theorem and the requirement that
$D_\mu \varphi_i   \rightarrow    0    $
as $r  \rightarrow    \infty$ then  imply    the presence of a magnetic flux
through the $(r,\theta)$ plane:
\begin{equation}
\oint_{S^1_{\infty}} W_{\theta} d\theta =2\pi n \, .
\end{equation}

As there are no
gauge orbits
   perpendicular to the $U(1)_{l}$ gauge orbits
 (because  $SU(2)_g$  is \emph{not}
 gauged),  any collective motion off such a  $U(1)_l$ orbit
(i.e., any unwinding of the string at infinity) would cost
an infinite amount of gradient energy\footnote{In a finite space,
this gradient energy barrier would be finite.}
which could not be compensated for by an
appropriate gauge connection.
Thus, it is
really the fundamental group of a
gauge orbit $S^1_{l}$ on the vacuum manifold
that matters for the stability against unwinding
and not the naive fundamental group of the vacuum manifold $S^3$  itself
\cite{VA1,Hindmarsh1}.

It is important to  note, however, that the stability against unwinding
(i.e., the conservation of the total magnetic flux as a topological invariant)
does not necessarily mean that such defects are stable against other
types of deformations. In particular, the magnetic flux need not be confined
or localized
in a small or  thin  region of spacetime, but may spread out to fill larger
regions (while still being  globally   conserved).
As was pointed out in \cite{Hindmarsh1,Hindmarsh2,Preskill},
topological defects may be unstable against such `spreading' if the full
vacuum
manifold is a non-trivial fiber bundle over a base manifold with  the gauge
orbit as the fiber such that the relevant homotopy group of the
total bundle becomes trivial.
To translate this to our situation, one notes that
 $S^3$ is a nontrivial $S^1_{l}$ bundle
over $\mathbb{C}\mathbb{P}^1\cong S^2$,
and $\pi_{1}(S^3)=\{1\}$, whereas $\pi_{1}(S^1_{l})
=\mathbb{Z}$.\footnote{In
\cite{Hindmarsh1} such a nontrivial fibering with a collapse
of the relevant homotopy group over the total bundle has been suggested as the
defining property of a semilocal defect. In \cite{Preskill}, the
defining property is chosen to be the ability to `spread' the magnetic flux.
}
 As we will see shortly, in
 our model  this phenomenon manifests itself in the form  of zero modes
that describe  a possible   fattening and warping
 of the magnetic flux tube. This ability to spread out
the flux is one of  the properties that make these semilocal strings
more appealing in a cosmological context, see the discussion in section 5.

Let us now describe the semilocal string solutions of our model in
more detail. The above-mentioned asymptotic behavior for $r
\rightarrow  \infty$ and the cylindrical symmetry imply that
$\Phi(r,\theta) \equiv (\varphi_1(r,\theta),\varphi_2 (r,\theta))$
approaches a field configuration of the form
\begin{equation}
\Phi(r,\theta) \longrightarrow \Phi_{0} e^{in\theta},
\qquad r\longrightarrow \infty
\end{equation}
where $\Phi^{\dagger}_{0}\Phi_{0}=\xi$.
 The global $SU(2)_g$ symmetry allows us to
align $\Phi_{0}$ with the $\varphi_1$ direction, and the most general
Higgs field configuration consistent with cylindrical symmetry is
then\footnote{Linear combinations such as  $\sum_{m}h^{(m)}(r)e^{im\theta}$
would break the assumed rotational symmetry,
as quantities such as the potential $V^D$ would in general no longer
 be $\theta$-invariant.}

\begin{equation}
\left(\begin{array}{c}
\varphi_{1}(r,\theta)\\
\varphi_{2}(r,\theta)
\end{array}\right) = \left( \begin{array}{c}
f(r)e^{in\theta}\\h(r) e^{im\theta}
\end{array} \right)
,
\end{equation}
whereas the gauge potential takes the form
\begin{eqnarray}
g W_\mu\, \rmd x^\mu = n\alpha (r) \,\rmd\theta \,
  \quad \rightarrow \quad
F=\ft12F_{\mu \nu }\,\rmd x^\mu \,\rmd x^\nu ={n \alpha '(r)\over
g} \rmd r\,\rmd\theta = {n \alpha '(r)\over  g C(r)} e^1 e^2 \,.
 \label{explicitW}
\end{eqnarray}

Here, $f(r)$, $h(r)$ and $\alpha(r)$ are  the profile functions
of the Higgs fields and the vector field, whereas
$m$ is an a priori  arbitrary integer.

The desired asymptotic behavior at $r \rightarrow    \infty$ and
the assumed non-singular behavior at the origin impose the
following boundary conditions on the functions $f(r)$, $h(r)$,
$\alpha(r)$ and the metric function $C(r)$:
\begin{eqnarray}
C(0)=0, &\qquad&
C^{\prime}(0)=1 \nonumber\\
f(0)=0, &\qquad& f(r)\longrightarrow \sqrt{\xi} \quad \textrm{ as }
r\longrightarrow \infty \nonumber\\
h(0)=h_{0}\delta_{m0}, &\qquad& h(r) \longrightarrow 0 \quad \,\,\,\,\,\,  \textrm{ as }
r \longrightarrow \infty\nonumber\\
\alpha(0)=0, &\qquad& \alpha(r) \longrightarrow 1 \quad
\,\,\,\,\,\,  \textrm{ as } r \longrightarrow \infty
\label{asymptotics}
\end{eqnarray}
where we have  indicated   that $h(0)$ might be non-zero for $m=0$, with
$h_{0}$ being  some constant.

Our goal now is to find profile functions $f(r)$, $h(r)$,  $\alpha(r)$,
$C(r)$, which
describe a solution of the
Lagrangian  that preserves $1/2 $ of the original supersymmetry.
We thus have to find an appropriate Killing spinor $\epsilon(r,\theta)$
such that $\delta \psi_{\mu}=\delta \chi_{i}= \delta \lambda=0$
can be satisfied for a certain   form of the profile functions.
For simplicity,  we will from now on   assume
\begin{equation}
n>0 \, .
\end{equation}
In order to cancel the magnetic flux contribution to $\delta \lambda$,
we choose  the following
projector for the Killing spinor
\begin{equation}
  \gamma ^{12}\epsilon\, = \, - \rmi \gamma _5 \epsilon\,.
 \label{projectioneps12}
\end{equation}

The BPS equations that follow from the vanishing of the transformation
rules~(\ref{susyf}) for  the chiral fields   $\chi_{i}$ and for the gaugino
$\lambda$ are then
\begin{eqnarray}
&&\Bigl( C(r){\partial}_r \, + \, {\rm i} \, ({\partial }_\theta-\rmi g W_{\theta} ) \Bigr)
\varphi_{i} \, = \, 0\,, \nonumber\\
 &&F_{12 } \, -   D\, = \, 0\,, \label{bpseqrtheta}
\end{eqnarray}
which leads  to
\begin{eqnarray}
    C (r)f'(r)&=& n\left[ 1-\alpha (r)\right] f(r) , \label{f}\\
 C (r)h'(r)&=& \left[ m-n\alpha (r)\right] h(r) , \label{h}\\
    \alpha '(r)&=& \frac{g^2 C(r)}{n}\left[\xi - f^2(r) -h^2(r)
    \right]. \label{alpha}
\end{eqnarray}
These three equations are not completely independent, as solving
eqs. (\ref{f}) and (\ref{h}) for $\alpha(r)$ and equating both
expressions yields the relation\footnote{In the special case $n=1$
and $m=0$, the flat space limit of  (\ref{lnhf}) reduces to the
expression $h(r)=(w/r)f(r)$ used in
 \cite{Hindmarsh1}. Analogues in conformally flat worldsheet coordinates
  also appear in \cite{GORS}.}
\begin{equation}
\partial_{r} \ln \left( \frac{h(r)}{f(r)} \right) = \frac{m-n}{C(r)}.\label{lnhf}
 \end{equation}
We make further use of this relation in Section 2.3 to derive an
upper bound on the possible integers $m$.

It remains to determine the gravitino BPS equation, for which one needs the
composite connection $A_{\mu}^B$:
\begin{equation} A_r^B=0\,,\qquad
 A_\theta ^B =\frac{1}{M_{P}^{2}} \left[   nf^2(r)+ mh^2(r) +n\alpha(r)
\left( \xi - f^2(r) -h^2(r) \right) \right]
.
 \label{explicitAB}
\end{equation}

The gravitino BPS equation is now
\begin{equation}
  \partial _r\epsilon =0\,,\qquad \left[ \partial _\theta +\ft12 iC'(r)
+\ft12\rmi A_\theta ^B\right] \epsilon _L(\theta)=0\,.
 \label{gravitinoBPS}
\end{equation}
The $\theta$ dependence of $\epsilon$ follows from the limit
$r \rightarrow    0$, where $A_{\theta}^{B}$  goes to zero
and     $C^{\prime}(r)$ approaches $1$
   so that (\ref{gravitinoBPS})
implies
\begin{equation}
  \epsilon _L(\theta) = \rme^{-\ft12\rmi\theta } \epsilon _{0L}\, ,
 \label{epsilontheta}
\end{equation}
where $\epsilon _{0L}$ is the left handed part of
  a constant spinor satisfying the projection
equation~(\ref{projectioneps12}). Thus, the gravitino equation
$\delta \psi_{\mu}=0$ boils down to
\begin{equation}
  1-C'(r) =  A_\theta ^B\,. \label{C}
 \label{diffeqrho}
\end{equation}
with $A^B_{\theta}$ as in (\ref{explicitAB}).

\subsection{Limiting cases}
While no analytic solution to the BPS equations (\ref{f}) --
(\ref{alpha}), (\ref{C}) is known, a great deal of physical
information can already be obtained by studying the limiting
behavior at $r=0$ and $r\rightarrow \infty$.

We will first show how the limiting behavior constrains the
possible integers $m$ in $\varphi_2 (r,\theta)$. In the limit where
$r$ approaches zero, $C(r)= r+\mathcal{O}(r^2)$ (see eqs.
(\ref{asymptotics})). As $\alpha(r)$ must be at least of linear
order in $r$, the $\alpha$-term in the BPS equations (\ref{f}) and
(\ref{h}) can be neglected to lowest order in $r$. To lowest
non-trivial
  order, these two equations thus read
\begin{eqnarray}
rf^{\prime}(r)&=&nf(r)\\
rh^{\prime}(r)&=&mh(r)
\end{eqnarray}
whence
\begin{equation}
f(r)=f_n r^n + \mathcal{O}(r^{n+1}), \qquad h(r) = h_{m} r^m + \mathcal{O}(r^{m+1})
\end{equation}
with  some  non-vanishing coefficients  $f_{n}$, $h_{m}$
($h_{m}$ might be zero for $m=0$).
Finiteness of $h(r)$  at $r=0$ then   implies $m\geq 0$.

To derive also an upper limit on $m$, consider  eq. (\ref{lnhf}).
As $h(r)/f(r) \rightarrow 0    $ for $r\rightarrow \infty$, the
left hand side of (\ref{lnhf}) has to be negative, which, because
of the positivity of $C(r)$ implies that $m<n$. We have thus
obtained the bounds
\begin{equation}
0\leq m < n .
\end{equation}

Let us now take a closer look at the limiting  behavior   at large
$r$. As $r$ approaches infinity, $(f(r),h(r),\alpha(r))$ approach
their asymptotic values $(\sqrt{\xi},0,1)$. It is easy to see,
that the four BPS-equations are consistent with this  behavior
provided that asymptotically
\begin{equation}
C(r)  \sim     r \left(1 - \frac{n\xi }{M_P^2}\right) \, .
\end{equation}
This shows that asymptotically the string creates a locally-flat
conical metric with an angular deficit angle
which is due to the constant FI term $\xi$:
\begin{equation}
\rmd s^2= -\rmd t^2 +\rmd z^2+\rmd r^2 + r^2\left(1-
\frac{n\xi}{M_P^2}\right)^2 \rmd \theta ^2\,.
 \label{metricFAR}
\end{equation}
Notice also that in   the  limit $r \rightarrow\infty$   the full
supersymmetry is restored because $F_{\mu\nu}=0$, $D=0$,
$\partial_r \varphi_i= D_\theta \varphi_i= 0$ and
$R_{\mu\nu}{}^{ab}=0$ which corresponds to  the enhancement of
supersymmetry away from the core of the string. This is the same
result as for the standard ANO vortex \cite{Dvali:2003zh},
however, it should be pointed out that the rate at which the
asymptotic values of the profile functions $f(r)$, $h(r)$ and
$\alpha(r)$ are obtained is in general no longer described by an
exponential fall-off, but can change into a   power law behavior
depending on the values of certain moduli parameters of  the
semilocal string \cite{Hindmarsh1} . To understand this better,
let us now  turn to    the behavior at small $r$.

Expanding the profile functions as  power series of the form
 $f(r)=f_0+f_1r+f_2r^2+\ldots$ and similarly
for $h(r)$, $\alpha(r)$ and $C(r)$,
 we know already that
\begin{eqnarray}
f(r)&=&f_n r^n +\mathcal{O}(r^{n+1})\\
h(r)&=&h_m r^m +\mathcal{O}(r^{m+1})\\
C(r)&=&r+\mathcal{O}(r^2)
\end{eqnarray}
In order to determine the lowest non-trivial power of $\alpha(r)$,
we expand (\ref{alpha}) to lowest order and obtain
\begin{equation}
\alpha(r)=\alpha_2 r^2 + \mathcal{O}(r^3)  , \qquad \textrm{ with } \qquad
\alpha_2=\frac{g^2}{2n}[\xi-h_{0}^2] \, \label{alphaorder1}.
\end{equation}

In a similar fashion one can now determine the higher coefficients
order by order and arrives at recursion relations that determine
all coefficients in terms of $\xi$ and the lowest non-trivial
coefficients $f_n$ and $h_m$ of the Higgs fields. It is easy to
see that the recursion relations are such that $f(r)$ contains
only odd powers when $n$ is odd and only even powers when $n$ is
even.  Similarly, the function $h(r)$ is odd/even if $m$ is
odd/even. $\alpha(r)$, by contrast is always even, whereas $C(r)$
only contains odd powers.

This power series expansion leaves the lowest Higgs coefficients
$f_n$ and $h_m$ undetermined. In fact, the coefficient $h_{m}$
turns out to be a undetermined parameter of the solution, whereas $f_n$
can be fixed numerically by matching to   the required asymptotic
behavior at infinity. The numerical value of $f_{n}$  depends upon
the parameter $h_{m}$.

The parameters $h_m$ can actually be taken to be complex, as one
has the freedom to shift the phase of $\varphi_2 =h(r)e^{im\theta}$
by a constant shift and absorb that phase into $h_{m}$. If one
breaks the rotational invariance by either allowing linear
combinations
$\varphi_2(r,\theta)=\sum_{m=0}^{n-1}h^{(m)}(r)e^{im\theta}$ and/or
by splitting an $n$-vortex into (up to $n$)  vortices with smaller
winding numbers and different positions in the $(r,\theta)$ plane,
one   expects   supersymmetry to be preserved  and finally arrives at a
$2n$-complex-dimensional moduli space of semilocal vortices
\cite{GORS}.

The physical interpretation of the moduli $h_m$ can most easily  be
illustrated   for the parameter $h_0$ in   the case   $m=0$.
In that case, eqs. (\ref{alphaorder1}) and  (\ref{explicitW})
yield the magnetic field
\begin{equation}
F_{12}(0)=g(\xi-h_0^2)
\end{equation}
at the origin. In the limit $h_0 \rightarrow 0$ this approaches
the magnetic field in the core of the ANO-string \cite{Dvali:2003zh}. This
is not surprising, as $h_0=0$ implies $h(r)\equiv 0$ in the case
of $m=0$ via the recursion relations that follow from the BPS
equations. In that case, the BPS equations reduce to those of the
ANO-string.

On the other hand, increasing $h_0$ from $0$ to $\sqrt{\xi}$,
decreases the magnetic field in the core. As the total flux only depends
on the winding number $n$, and thus is unaffected by changes in $h_{0}$,
 this means that the flux tube must become flatter and wider.
Thus, $h_0$ is a measure for  the width of the magnetic flux tube
and encodes the possibility for semilocal strings to spread out
their flux into larger  space-time regions \cite{Hindmarsh1}.

Another interesting quantity is the spacetime curvature
$R=2C^{\prime\prime}/C$
in the center of the string. At $r=0$, this quantity reduces to
$R(0)=12C_{3}$. $C_3$ can be easily obtained from the first  non-trivial power
of eq. (\ref{C}), which leads to
\begin{equation}
R(0)=-4  \left(\frac{f_{1}}{M_{P}} \right)^{2} -
2\left(\frac{g(\xi-h_{0}^2)}{M_{P}} \right)^2
\end{equation}

We finally note that for  high $n$ and $m$, the functions $f(r)$ and $h(r)$
are zero at least   up to order $r^{m}$  (remembering $m<n$).
Up to terms of order $r^{2m}$, the BPS equations (\ref{alpha}) and (\ref{C})
then simplify to
\begin{equation}
\alpha ' (r) = \frac{g^2 \xi}{n} C (r)\,,\qquad 1-C'(r) - n
M_P^{-2}\xi \alpha(r) =0\,, \label{limitf0}
\end{equation}
which, in view of the boundary conditions at $r=0$, can be solved   as
\begin{eqnarray}
\alpha (r) & = &  \frac{M_P^2}{n\xi }\Bigl(1-\cos( {g\xi\over M_P} r)\Bigr), \nonumber\\
C (r) & = & {M_P\over \xi g} \sin ( {g\xi\over M_P} r) \, .
\label{solvef0a}
\end{eqnarray}
Up to terms of order $r^{2m+1}$, the metric at small $r$ is then:
\begin{equation}
\rmd s^2= -\rmd t^2 +\rmd z^2+\rmd r^2 + \left[{M_P\over \xi g}
\sin ( {g\xi\over M_P} r)\right]^2 \rmd \theta ^2\,,
 \label{metricCORE}
\end{equation}
whereas, up to terms of order $r^m$ for  the Higgs and  $r^{2m+2}$
for the   gauge   fields, one obtains
\begin{equation}
\varphi_i= 0\, , \qquad  W_\theta= \frac{M_P^2}{g\xi }\Bigl(1-\cos(
{g\xi\over M_P} r)\Bigr) \,, \qquad F_{r\theta}= M_P \sin ( {g\xi\over
M_P} r)\,.
\end{equation}

\section{Cosmology and semilocal strings}

\subsection{Inflation}

The semilocal strings embedded into supergravity were shown in
the previous section to have an unbroken supersymmetry for a continuous range
of the
parameter $h_0$, which effectively measures the width of the flux
tube and is not defined by the parameters in the action. Now we
would like to study the properties of the semilocal strings in
applications to cosmology.

The generalized D-term model after doubling the number of
hypermultiplets corresponding to doubling the number of  $D7$ branes
in the $D3/D7$ model has the following potential:
\bea
V&=&\frac{g^2}{2}\left|\phi_+\phi_--\tilde\phi_-\tilde\phi_+\right|^2+
\frac{g^2}{8}\left[|\phi_+|^2+|\tilde\phi_+|^2-|\phi_-|^2-|\tilde\phi_-|^2-2\xi
\right]^2+ \nonumber\\ &
&\frac{g^2}{2}|s|^2\left[|\phi_+|^2+|\phi_-|^2+|\tilde\phi_+|^2+|\tilde\phi_-|^2\right].
\label{potphi} \eea
Here all scalar fields have canonical kinetic terms, such as
${1\over 2} (\partial \phi)^2$. They are related to the chiral
fields in the action (\ref{potphi}) as follows: $\phi_+ = \sqrt 2
\, \Phi_+, s= \sqrt 2 \, S$ etc. The tree-level potential
(\ref{potphi}) has a supersymmetric minimum (a valley with $V=
0$):
\bea &
&s=0\ , \qquad\left|\phi_+ \phi_- -
\tilde\phi_+ \tilde\phi_-\right| \ = \ 0 \ , \nonumber\\
& &|\phi_+|^2 + |\tilde\phi_+|^2 - |\phi_-|^2 - |\tilde\phi_-|^2 \
= \ 2\xi  \label{v1} \ . \eea The $U(1)$  gauge   symmetry is broken
in the minimum. In
the absence of strings, the fields can take any values along this
valley. When strings are present, the fields tend to drift towards
the state with $\phi_-=\tilde\phi_-=0$, because
$\phi_-=\tilde\phi_-=0$ for the BPS (lowest energy state)
solutions \cite{PRTT96,ADPU02}.

The potential has a second flat direction, $\phi_\pm =
\tilde\phi_\pm =0$. Inflation occurs when the field $s$ rolls
along this flat direction. At the classical level, the scalar
potential (\ref{potphi}) equals $\frac{g^2}{2}\xi^2$ for all
values of $s$ at $\phi_\pm = \tilde\phi_\pm =0$; in order to study
inflationary dynamics one should take into account one-loop
corrections to the effective potential.

We will consider the models with $g^2 \gtrsim 10^{-4}$, because it
is more difficult to motivate much smaller coupling constant $g^2$
in string theory.  As we will see, inflation in the model with
$g^2 \gtrsim 10^{-4}$ ends at $s \gg s_c=\sqrt\xi$. In this
regime, the one-loop corrected  potential of the field $s$ in the
model with $C$ hypermultiplets of equal $g^2$ is given by
\cite{pterm,UAD}
\be V_{\rm
eff}=\frac{g^2}{2}\xi^2\left\{1+\frac{Cg^2}{8\pi^2}\left[\ln\frac{s^2}{s_c^2}\right]\right\}
\ .\ee
To study inflation, one should use the Friedmann equation
\begin{equation}\label{infl}
H = {\dot a \over a} \approx \sqrt{V/3} \approx {g\xi\over \sqrt
6} \ ,
\end{equation}
where $a(t)$ is a scale factor of the universe. This leads to
inflation
\begin{equation}\label{infl1}
a(t) = a(0)~ \exp{ {{g \xi\  t \over\sqrt 6} }}~ .
\end{equation}
During the slow-roll regime, the field $s$ obeys equation $3H\dot s
= -V'(s)$ \cite{lindebook}, which gives
\begin{equation}\label{infl2}
s^2(t) = s^2(0) - {Cg^3\xi ~t\over 2\sqrt 6 \pi^2} \ .
\end{equation}

Using equations (\ref{infl1}), (\ref{infl2}) one can find the
value of the field $s_{{}_N}$ such that the universe inflates
$e^N$ times when the field rolls from $s_{{}_N}$ to the
bifurcation point $s = s_c$:
\begin{equation}\label{infl3}
s_{{}_N}^2  = {s_c}^2  + {Cg^2 N\over 2 \pi^2} = {\xi}  + {Cg^2
N\over 2 \pi^2} \ .
\end{equation}
In this paper we will concentrate on the case ${Cg^2\over 2
\pi^2}> \xi$. Density perturbations on the scale of the  present
cosmological horizon have been produced at $s \sim s_{N} \gg s_c$ with $N
\sim 60$, and their amplitude is proportional to ${V^{3/2}\over V'
}$ at that time \cite{lindebook}. One can find $\xi$ using the COBE
normalization for inflationary perturbations of the metric on the
horizon scale \cite{Lyth:1998xn}:
\begin{equation}\label{smallg}
\delta_H ={1\over 5\sqrt 3 \pi} {V^{3/2}\over V'  }\approx
1.9\times 10^{-5} \ .
\end{equation}
This yields
\begin{equation}\label{smallga}
 {V^{3/2}\over V'  }=  {2\sqrt 2\pi^2 \xi\over C g}~ s_{{}_N}\sim 2\pi \xi\sqrt{N\over C}  \sim
5.2 \times 10^{-4} \ .
\end{equation}
In what follows, we will concentrate on the case with two
hypermultiplets and take $N = 60$. In this case
\begin{equation}\label{ksi}
\xi \sim {5.2 \times 10^{-4}\over 2\pi} \sqrt{C/N} \sim 1.5 \times
10^{-5} \ .
\end{equation}
This yields the scale of spontaneous symmetry breaking
\begin{equation}\label{ksi2}
\sqrt\xi \sim 9.3 \times 10^{15}\, {\rm GeV}.
\end{equation}

This model has a nearly flat spectrum of density perturbations, $n
\approx 0.98$, and a very small amplitude of gravitational waves.
Both of these features are in a good agreement with the
observational data.

\subsection{Semilocal strings in cosmology}

After the field $s$ passes the bifurcation point $s_c = \sqrt\xi$, the fields
$\phi_\pm,\tilde\phi_\pm = 0$ start rolling to the minimum of the effective
potential at $|\phi_+|^2 + |\tilde\phi_+|^2 - |\phi_-|^2 -
|\tilde\phi_-|^2 \ = \ 2\xi$. This leads to production of cosmic
strings.  One should check whether the semilocal cosmic strings
produced during this process can affect the amplitude of
large-scale density perturbations.

Cosmic strings lead to perturbations of the metric with flat spectrum
if in the course of the evolution of the universe there always
remains at least one string of length comparable to the size of
the particle horizon $\sim ct$ \cite{VilenkinShellard}. We will
call such strings infinite. Thus one needs to check whether
infinite semilocal strings can form and survive long after the end
of inflation.

This issue is rather nontrivial. Even though stability of
semilocal strings is not protected by topological considerations,
one could argue that infinitely long straight strings are
protected because they are BPS solutions, so their energy cannot
get any lower. On the other hand, the fact that the thin and thick
BPS strings have the same tension indicates that strings can be
unstable if it is possible to excite a zero mode which would lead
to the growth of their thickness; see e.g. \cite{Hindmarsh1,Hindmarsh2,GORS}. However,
it would require infinite action to excite such a zero mode in an
infinitely large universe \cite{GORS}. Also, the growth of
thickness of a single infinite string occurs with the speed
smaller than the speed of light, and the string tension does not
change in this process. Meanwhile (however strange this might
seem), the diameter of the observable part of the universe grows
with the speed 4 times greater than the speed of light in a
universe dominated by ultrarelativistic particles, and with the
speed 6 times greater than the speed of light in the cold dark
matter dominated universe \cite{lindebook}. Thus, on the horizon
scale, each {\it separate} infinite semilocal string will always look
thin, and its expansion will not lead to the change of its
tension. This means that the spreading of each individual BPS
string would not change the amplitude of density
perturbations which it produces.

However, the strings created at the end of inflation initially are
not straight, infinite BPS strings. In contrast to the case of the
standard Nielsen-Olesen cosmic strings, infinite semilocal strings
do not appear at the moment of the phase transition. Typically,
soon after the phase transition, strings look like a cloud of string
segments of a size comparable to the correlation length at the
moment of the cosmological phase transition. These  string
segments have global monopoles at their ends. Each global monopole
attracts a nearby global anti-monopole with a force that does not
depend on the distance between them. If the monopole and the
antimonopole closest to it belong to the ends of two different
strings, they attract each other, and the two strings tend to
merge, forming one longer string. However, if a string is shorter
than the distance between it and another string, the two monopoles
on its ends are attracted to each other stronger than to other
monopoles. As a result, either  such short string segment shrinks
and disappears due to its tension and monopole interactions, or in the beginning it
forms a small loop, and then the loop shrinks and disappears.

Infinite semilocal strings  responsible for the large scale
density perturbations can appear only because of merging of short
string segments. The probability of this process is expected to be
small; it is not reinforced by topological considerations. In
addition, the possibility of growth of thickness of the near-BPS
string segments eventually may lead to their overlapping with
other string segments, which may result in their de-localization
and effective disappearance. (Note that it does not cost infinite
energy or infinite action to excite the zero mode for each segment
of a semilocal near-BPS string of a finite size.) A numerical
investigation of this process was performed in \cite{ABL99}, with
the conclusion that infinite semilocal BPS strings are not formed,
and therefore they do not lead to the problem of excessively large
string-related metric perturbations.

One should exercise some caution before applying the results  of Ref.
\cite{ABL99} to the  model (\ref{potphi}). Indeed,  numerical simulations
performed in Ref. \cite{ABL99} did not fully take into account the quantum
dynamics of spontaneous symmetry breaking in hybrid inflation. Instead of that,
the authors simply made some plausible assumptions about the random
distribution of the scalar field after the phase transition. Then they verified
that their results are not very sensitive to these assumptions, but in their
analysis they did not take into account the expansion of the universe. Also,
the numerical analysis  was performed in \cite{ABL99} for a simpler model,
which did not contain the fields $\phi_-,\tilde\phi_-$. One could argue
\cite{UAD} that these fields are not important because  eventually the fields
drift towards the state with $\phi_-=\tilde\phi_-=0$ since
$\phi_-=\tilde\phi_-=0$ for the BPS solutions \cite{PRTT96,ADPU02}. The
short string segments produced after inflation initially are not in the BPS
state, and the process of disappearance of the fields $\phi_-=\tilde\phi_-=0$
could take a very long time \cite{PRTT96}, but the numerical simulations
performed in \cite{Pickles:2002ym} suggest that this time can be rather short.

We do believe that the results obtained in \cite{ABL99, Pickles:2002ym}
provide a strong evidence that the semilocal strings formed in the model
(\ref{potphi}) do not pose any cosmological problems, but the importance of
this subject warrants its full quantum mechanical analysis. This can be done
using the methods of  Ref. \cite{Felder:2000hj}, where the process of
spontaneous symmetry breaking in the simplest versions of hybrid inflation was
investigated. These methods allow to perform a complete quantum field theory
investigation of string formation in the early universe \cite{Tkachev:1998dc}.
Leaving this full analysis for a future investigation, we will discuss some
important features of the processes which occur after inflation in the model
(\ref{potphi}). We will argue that the expansion of the universe does not lead
to qualitative changes in the processes studied in \cite{ABL99}, and that in
the first approximation one can indeed ignore the effects related to the fields
$\phi_-,\tilde\phi_-$. Our analysis will provide a framework for the future
numerical investigation of this issue.

\subsection{Symmetry breaking and string formation at the end of
inflation}

According to \cite{pterm,UAD}, the field $s$ at the last stages of
inflation rolls down along the valley with
$\phi_\pm,\tilde\phi_\pm = 0$ until it reaches the bifurcation
point $s_c = \sqrt\xi$. Then it continues rolling along the ridge
with $\phi_\pm,\tilde\phi_\pm = 0$ until  the fields
fall down from the ridge and spontaneous symmetry breaking occurs.
When the fields $\phi_\pm,\tilde\phi_\pm$ acquire nonvanishing
(complex) expectation values, their initial distribution looks
very chaotic, which can lead to string formation via the Kibble
mechanism.

Even though the discussion of spontaneous symmetry breaking can be
found in every book of quantum field theory, the dynamics of this
process was understood only very recently, and the results were
rather surprising \cite{Felder:2000hj}. For example, one could
expect that when the fields fall down from the top of the
effective potential to its minimum at $\phi \sim \sqrt{2\xi}$, they
should enter a long stage of oscillations with the amplitude $\sim
\sqrt {2\xi}$. However, a detailed investigation of this issue shows that because of the exponential growth of
quantum fluctuations in the process of spontaneous symmetry
breaking,  the main part of the kinetic energy of the rolling
fields becomes converted into the gradient energy of the inhomogeneous distribution of these
fields within a single oscillation. The resulting distribution looks like a collection of classical waves of scalar fields moving near the minimum of the effective potential, the amplitude of these waves being much smaller than the amplitude of spontaneous symmetry breaking  $\sqrt {2\xi}$. In other words, the
main part of the process of symmetry breaking completes within a
single oscillation of the field distribution \cite{Felder:2000hj}.

In order to study this process in the context of the  model
(\ref{potphi}), we need to know the speed of rolling of the field
$s$ as it passes the bifurcation point $s_c$. This will allow us
to find how far along the ridge the field $s$ can go
until spontaneous symmetry breaking occurs. The estimates we are
going to make will be very rough, but sufficient for our purposes.

According to (\ref{infl3}) and  (\ref{ksi}), for $C= 2$ and $g^2 \gg 10^{-4}$ the slow-roll
regime ends not at the bifurcation point, but earlier,  at
$s_1 \sim  {g \over \pi } \gg s_c$. At $s < s_1$ the field rolls down very fast,
reaching the bifurcation point $s_c$ within the time smaller than
the Hubble time $H^{-1}$. Therefore during this time the total
energy density of the field $s$ is (approximately) conserved. This
leads to an estimate ${1\over 2} \dot s^2 \sim {g^4\xi^2\over
16\pi^2} \ln{g^2\over \pi^2 \xi}$ for the velocity of the field
$s$ at the bifurcation point.

Symmetry breaking happens at $s < s_c$ because the fields
$\phi_+,\tilde\phi_+$ acquire the tachyonic mass at $s < s_c$,
\be m_+^2 = g^2(s^2-s_c^2) <0 \ . \ee
This means that, after passing the bifurcation point, the tachyonic
mass of the fields $\phi_+,\tilde\phi_+$ grows as follows:
\be m_+^2 \approx  -2g^2\sqrt \xi \Delta s = g^2\sqrt \xi\, \dot
s\, \Delta t \ ,  \ee
where $\Delta s = s_c - s$.

The process of spontaneous symmetry
breaking occurs because of the exponential growth of quantum
fluctuations of the fields with a tachyonic mass. Ignoring, in the
first approximation, the time dependence of the tachyonic mass,
one can find that quantum fluctuations of the fields with $
m_+^2  <0$ and momentum $k < |m_+|$ grow exponentially:
\be \delta \phi_+ \sim \delta \tilde\phi_+ \sim e^{i\sqrt
{m_+^2+k^2}\ t} \sim e^{\sqrt {|m_+^2|- k^2}\ t} \ee
Occupation numbers of quantum fluctuations grow as $e^{2\sqrt {|m_+^2|- k^2}\ t}$. When the occupation numbers become much greater than 1, growing quantum fluctuations can be interpreted as a classical scalar field.   This means that the growing components of the scalar fields $\phi_+,
\tilde\phi_+$ will look inhomogeneous on the scale $|m_+|^{-1}$.
The average initial amplitude of quantum fluctuations with $k <
|m_+|$ is $O(10^{-1}|m_+|)$  \cite{Felder:2000hj}. Because of the
exponential character of the growth, it takes the field
distribution the time $\sim  |m_+|^{-1} \ln {10\sqrt{\xi}\over
|m_+|}$ before it reaches the minimum of the effective potential.
In fact, the total duration will be somewhat smaller because the
curvature of the effective potential grows in the process.

Note that in our case $|m_+|$ is vanishingly small at the
bifurcation point; it becomes large only later. Therefore
spontaneous symmetry breaking occurs only when the time that it
takes for the rolling field $s$ to reach some point $s < s_c$
becomes comparable to $|m_+(s)|^{-1} \ln {10\sqrt{\xi}\over |m_+(s)|}$.
This condition reads
\be 2 g^2 \sqrt\xi\ \Delta s \ln^{-2} {10\sqrt\xi\over |m_+(s)|}
\sim {\dot s^2 \over (\Delta s)^2} \sim { g^4\xi^2 \ln{g^2\over
\pi^2 \xi} \over 8\pi^2(\Delta s)^2}. \ee
To find an approximate solution of this equation, we will take
$|m_+(s)| = |m_+(0)| = g\sqrt \xi$ under the logarithm. This
yields
\be \Delta s   \sim F(g) \  g^{2/3}  \sqrt\xi\ ,  \ee
where
\be F(g)=  \left({\ln{g^2\over 4\pi \xi}\ln^2{10^2\over g^2}\over
64\pi^2 }\right)^{1/3} . \ee
Numerical evaluation of this expression shows, for example, that
$F \sim 0.8$,   and $\Delta s \sim 0.2\, s_c$   for $g^2\sim 10^{-2}$.
Finally, we find the absolute value of the tachyonic mass of the
fields $\phi_+,\tilde\phi_+$ at the moment when the symmetry
breaking occurs:
\be |m_+|   \sim  2^{1/4} F^{1/2}(g) \ g^{4/3} \sqrt\xi\ =
O(g^{4/3} \sqrt\xi) \ . \ee

Note that at this stage the only fluctuations experiencing the
tachyonic growth are those of the fields $\phi_+,\tilde\phi_+$;
the fields $\phi_-,\tilde\phi_-$ are not generated. On average,
the amplitude of the complex fields $\phi_+$ and $\tilde\phi_+$
should be the same, but in different parts of the universe
separated by a distance more than $m_+^{-1}$ the values of these
fields as well as their phases can be different. Therefore the
correlation length which determines  the typical initial length of the string segments and the typical initial distance between the cosmic strings is expected to be \be \Delta x =
O(g^{4/3} \sqrt\xi)\ . \ee For the range of parameters we are
interested in, this correlation length can be two orders of
magnitude smaller than the horizon size at that time,  $H^{-1}
\sim g^{-1}\, \xi^{-1}$, which means that the corresponding momenta are two orders of
magnitude greater than $H$.

This implies, in particular, that one can neglect the expansion of the
universe at the first stage of string formation. This removes one
of the concerns related to the applicability of the results of
Ref. \cite{ABL99} to the investigation of string formation after
hybrid inflation.

The simple tachyonic description of the process of spontaneous
symmetry breaking is valid only at the first stages of the
process. The effective potential of the fields $s$ and
$\phi_+,\tilde\phi_+$ far away from the ridge $\phi_+,\tilde\phi_+
= 0$ is very complicated, so one could expect that the fields will
move down along a very complicated trajectory. Rather
surprisingly,  the field $s$ and the (quasi)homogeneous component
of the fields $\phi_+,\tilde\phi_+$ follow a linear trajectory.
Once the amplitude of the fields $\phi_+,\tilde\phi_+$ grows up to
$O(\Delta s)$, they start falling down along the straight line
 \bea \label{line}
 \tilde\phi_+ &=& f\, \phi_+ \ ,
 \nonumber\\
 \sqrt{1+f^2\over 2}\, \phi_+ \, +\, s  &=&  \sqrt{\xi}\ \ ,
  \eea
where $f$ is an arbitrary constant depending on the local value of
the ratio between $\phi_+$ and $\tilde\phi_+$ when they approach
the linear trajectory. A similar regime for the nearly homogeneous component of the scalar fields was found in
\cite{Bastero-Gil:1999fz} in a model containing only two
time-dependent  fields, $\phi_+$ and $s$.

Eventually the fields $s$, $\phi_+$, $\tilde\phi_+$ roll to the
minimum of the potential at $s = 0$, $\phi_+^2 + \tilde\phi_+^2 =
2\xi$ and start oscillating there. When these fields reach the
minimum and continue moving forward by inertia, a certain
combination of the fields $\phi_-, \tilde\phi_-$ acquires a
tachyonic mass, and the corresponding fluctuations begin growing
exponentially. However, this happens only during a very short
period of time, of the same order as the inverse tachyonic mass of
the fields $\phi_-, \tilde\phi_-$ (without the important logarithmic factor
$\ln {10\sqrt{\xi}\over |m_+|}$, which appears in the description
of the growth of the fluctuations of the fields $\phi_-,
\tilde\phi_-$). As a result, the amplitude of the quantum
fluctuations of the fields $\phi_-, \tilde\phi_-$ does not grow
much and remain much smaller than the amplitude of the fields
$\phi_+, \tilde\phi_+ \sim \sqrt \xi$.

One should note that this description is a bit
oversimplified. Strictly speaking, the solution (\ref{line}),  is
valid only for homogeneous fields.  In our case, the homogeneity
is broken by tachyonic fluctuations on the scale $(g^{4/3}\,
\sqrt\xi)^{-1}$, but one can expect that, in the first
approximation, Eq. (\ref{line}) should still hold in each domain
of size $(g^{4/3}\, \sqrt\xi)^{-1}$ because the curvature of the
effective potential along the main part of this trajectory is
higher than $(g^{4/3}\, \sqrt\xi)^{2}$. On a larger
scale, the distribution of the fields $\phi_+,\tilde\phi_+$ is
very inhomogeneous. At the time when the fields reach the minimum
of the effective potential, the field distribution  looks like a
  collection of classical waves of all scalar fields. These waves
may move, collide, and produce waves of all other fields,
including the fields $\phi_-,\tilde\phi_-$. However, as we already
mentioned, a numerical investigation of similar models in
\cite{Felder:2000hj} shows that the amplitude of the waves
generated during this process is significantly smaller than the
amplitude of the fields $\phi_+, \tilde\phi_+ = O(\sqrt \xi)$.

This means, in particular, that in the first approximation one can
neglect the role of the fields $\phi_-,\tilde\phi_-$ in the
process of symmetry breaking and string formation. This is an
additional argument suggesting that the results of the investigation
of string formation in Ref. \cite{ABL99} in the model not
involving the fields $\phi_-,\tilde\phi_-$ should remain valid for
our model as well.

One should emphasize that the process described above is extremely
complicated. Therefore it would be important to verify our
analysis  by performing a full numerical investigation of symmetry
breaking in this model by the methods developed in
\cite{Felder:2000hj}. We hope to return to this issue in a
subsequent publication. However, on the basis of our analysis we
do expect that the main argument contained in \cite{UAD} and in
our work should be valid: By adding a new hypermultipet to the
inflationary model discussed above, one can avoid the excessive
production of metric perturbations related to cosmic strings,
while preserving the usual inflationary perturbations.

\section*{Acknowledgments}
It is a pleasure to thank G. Dvali, S. Kachru, J. Polchinski, M. M.
Sheikh-Jabbari, and D. Tong for useful discussions. This work was
supported  by  NSF grant 0244728. The work of K.D. is also supported
by  David and Lucile Packard Foundation Fellowship 2000-13856.
J.H. is supported in part by NSF
Graduate Research Fellowship. M.Z. is supported by  an Emmy-Noether-Fellowship
of the German Research Foundation (DFG), grant number ZA 279/1-1.


\begin{thebibliography}{99}

\bibitem{Dvali:1998pa}
G.~R.~Dvali and S.~H.~H.~Tye,
``Brane inflation,''
Phys.\ Lett.\ B {\bf 450}, 72 (1999)
[arXiv:hep-ph/9812483];
S.~H.~S.~Alexander,
``Inflation from D - anti-D brane annihilation,''
Phys.\ Rev.\ D {\bf 65}, 023507 (2002)
[arXiv:hep-th/0105032];
C.~P.~Burgess, M.~Majumdar, D.~Nolte, F.~Quevedo, G.~Rajesh and R.~J.~Zhang,
``The inflationary brane-antibrane universe,''
JHEP {\bf 0107}, 047 (2001)
[arXiv:hep-th/0105204];
C.~Herdeiro, S.~Hirano and R.~Kallosh,
``String theory and hybrid inflation / acceleration,''
JHEP {\bf 0112}, 027 (2001)
[arXiv:hep-th/0110271];
C.~P.~Burgess, P.~Martineau, F.~Quevedo, G.~Rajesh and R.~J.~Zhang,
``Brane antibrane inflation in orbifold and orientifold models,''
JHEP {\bf 0203}, 052 (2002)
[arXiv:hep-th/0111025]
B.~s.~Kyae and Q.~Shafi,
``Branes and inflationary cosmology,''
Phys.\ Lett.\ B {\bf 526}, 379 (2002)
[arXiv:hep-ph/0111101];
J.~Garcia-Bellido, R.~Rabadan and F.~Zamora,
``Inflationary scenarios from branes at angles,''
JHEP {\bf 0201}, 036 (2002)
[arXiv:hep-th/0112147];
J.~H.~Brodie and D.~A.~Easson,
``Brane inflation and reheating,''
JCAP {\bf 0312}, 004 (2003)
[arXiv:hep-th/0301138].



\bibitem{DHHK}K.~Dasgupta, C.~Herdeiro, S.~Hirano and R.~Kallosh,
``D3/D7 inflationary model and
M-theory'', Phys.\ Rev.\ D65 (2002) 126002,
[arXiv:hep-th/0203019].
.

\bibitem{KKLMMT}
S.~Kachru, R.~Kallosh, A.~Linde, J.~Maldacena, L.~McAllister and
S.~P.~Trivedi, ``Towards inflation in string theory'', JCAP 0310
(2003) 013, [arXiv:hep-th/0308055].

\bibitem{KKLT}
S.~Kachru, R.~Kallosh, A.~Linde and S.~P.~Trivedi, ``De Sitter
vacua in string theory'', Phys.\ Rev.\ D68 (2003) 046005,
[arXiv:hep-th/0301240].

\bibitem{Burgess:2003ic}
C.~P.~Burgess, R.~Kallosh and F.~Quevedo,
``de Sitter string vacua from supersymmetric D-terms,''
JHEP {\bf 0310}, 056 (2003)
[arXiv:hep-th/0309187].

\bibitem{Denef:2004dm}
F.~Denef, M.~R.~Douglas and B.~Florea,
``Building a better racetrack,''
[arXiv:hep-th/0404257].

\bibitem{DeWolfe:2004qx}
O.~DeWolfe, S.~Kachru and H.~Verlinde,
``The giant inflaton,''
[arXiv:hep-th/0403123];
 N.~Iizuka and S.~P.~Trivedi, ``An
inflationary model in string theory,'' [arXiv:hep-th/0403203];
A.~Buchel and A.~Ghodsi,
``Braneworld inflation,''
[arXiv:hep-th/0404151];
M.~Berg, M.~Haack and B.~K\"ors,
``Loop corrections to volume moduli and inflation in string theory,''
[arXiv:hep-th/0404087].


\bibitem{Silverstein:2003hf}
E.~Silverstein and D.~Tong,
``Scalar speed limits and cosmology: Acceleration from D-cceleration,''
[arXiv:hep-th/0310221];
M.~Alishahiha, E.~Silverstein and D.~Tong,
``DBI in the sky,''
[arXiv:hep-th/0404084].

\bibitem{Hsu:2004hi}
J.~P.~Hsu and R.~Kallosh,
``Volume stabilization and the origin of the inflaton shift symmetry in string
theory,''
JHEP {\bf 0404}, 042 (2004)
[arXiv:hep-th/0402047].

\bibitem{KalloshHsuProk}
J.~P.~Hsu, R.~Kallosh and S.~Prokushkin,
``On brane inflation with volume stabilization,''
JCAP {\bf 0312}, 009 (2003)
[arXiv:hep-th/0311077].

\bibitem{KTW}
F.~Koyama, Y.~Tachikawa and T.~Watari, ``Supergravity analysis of
hybrid inflation model from D3-D7 system'', [arXiv:hep-th/0311191];
H.~Firouzjahi and S.~H.~H.~Tye,
``Closer towards inflation in string theory,''
[arXiv:hep-th/0312020].

\bibitem{Kallosh:2004rs}
R.~Kallosh and S.~Prokushkin,
``Supercosmology,''
[arXiv:hep-th/0403060].




\bibitem{Angelantonj:2003zx}
C.~Angelantonj, R.~D'Auria, S.~Ferrara and M.~Trigiante,
 ``K3 x T**2/Z(2) orientifolds with fluxes, open string moduli  and critical
points,'' [arXiv:hep-th/0312019].


\bibitem{L}
A.~D.~Linde, ``Hybrid Inflation,'' Phys.\ Rev.\ D {\bf 49}, 748
(1994) [arXiv:astro-ph/9307002].

\bibitem{Binetruy:1996xj}
P.~Bin\'{e}truy and G.~R.~Dvali,
``D-term inflation,''
Phys.\ Lett.\ B {\bf 388}, 241 (1996) [arXiv:hep-ph/9606342];
E.~Halyo,
``Hybrid inflation from supergravity D-terms,''
Phys.\ Lett.\ B {\bf 387}, 43 (1996) [arXiv:hep-ph/9606423].

\bibitem{BDKV} P.~Bin\'{e}truy, G.~Dvali, R.~Kallosh and  A.~Van Proeyen,
``Fayet-Iliopoulos Terms in Supergravity and Cosmology'',
[arXiv:hep-th/0402046].

\bibitem{pterm}
R.~Kallosh and A.~Linde,
``P-term, D-term and F-term inflation,''
JCAP {\bf 0310}, 008 (2003), [arXiv:hep-th/0306058].



\bibitem{Avelino:2003nn}
P.~P.~Avelino and A.~R.~Liddle,
``Cosmological perturbations and the reionization epoch,''
Mon.\ Not.\ Roy.\ Astron.\ Soc.\  {\bf 348}, 105 (2004)
[arXiv:astro-ph/0305357].

\bibitem{Pogosian:2004mi}
L.~Pogosian and A.~Vilenkin, ``Early reionization by cosmic strings
revisited,'' [arXiv:astro-ph/0405606].

\bibitem{Kofman:2002rh}
L.~Kofman and A.~Linde,
``Problems with tachyon inflation,''
JHEP {\bf 0207}, 004 (2002)
[arXiv:hep-th/0205121].


\bibitem{Jones:2003da}
N.~T.~Jones, H.~Stoica and S.~H.~H.~Tye,
``The production, spectrum and evolution of cosmic strings in brane
inflation,''
Phys.\ Lett.\ B {\bf 563}, 6 (2003)
[arXiv:hep-th/0303269].

\bibitem{Dvali:2003zj}
G.~Dvali and A.~Vilenkin,
``Formation and evolution of cosmic D-strings,''
JCAP {\bf 0403}, 010 (2004)
[arXiv:hep-th/0312007].

\bibitem{Dvali:2003zh}
G.~Dvali, R.~Kallosh and A.~Van Proeyen,
``D-term strings,''
JHEP {\bf 0401}, 035 (2004)
[arXiv:hep-th/0312005].

\bibitem{Halyo:2003uu}
E.~Halyo,
``Cosmic D-term strings as wrapped D3 branes,''
JHEP {\bf 0403}, 047 (2004)
[arXiv:hep-th/0312268].

\bibitem{Copeland:2003bj}
E.~J.~Copeland, R.~C.~Myers and J.~Polchinski,
``Cosmic F- and D-strings,''
[arXiv:hep-th/0312067];
M.~G.~Jackson, N.~T.~Jones, J.~Polchinski, ``Collisions of Cosmic F- and D-strings,''
[arXiv:hep-th/0405229]

\bibitem{Lazar}
G.~Lazarides and C.~Panagiotakopoulos,
``Smooth hybrid inflation,''
Phys.\ Rev.\ D {\bf 52}, 559 (1995)
[arXiv:hep-ph/9506325];
R.~Jeannerot, S.~Khalil, G.~Lazarides and Q.~Shafi, ``Inflation
and monopoles in supersymmetric SU(4)c x SU(2)L x SU(2)R,'' JHEP
{\bf 0010}, 012 (2000) [arXiv:hep-ph/0002151].

\bibitem{Lyth:1998xn}
D.~H.~Lyth and A.~Riotto,
``Particle physics models of inflation and the
cosmological density  perturbation,''
Phys.\ Rept.\  {\bf 314}, 1 (1999)
[arXiv:hep-ph/9807278].

\bibitem{K01}
R.~Kallosh,
``N = 2 supersymmetry and de Sitter space,''
[arXiv:hep-th/0109168].


\bibitem{Endo:2003fr}
M.~Endo, M.~Kawasaki and T.~Moroi,
``Cosmic string from D-term inflation and curvaton,''
Phys.\ Lett.\ B {\bf 569}, 73 (2003)
[arXiv:hep-ph/0304126].



\bibitem{UAD}
J.~Urrestilla, A.~Ach\'{u}carro and A.C.~Davis,
``D-term inflation without cosmic strings,'' [arXiv:hep-th/0402032].



\bibitem{VA1}
T.~Vachaspati and A.~Ach\'{u}carro, {``Semilocal cosmic strings''},
Phys.Rev.  {\bf D44},  3067 (1991).

\bibitem{Hindmarsh1}
M.~Hindmarsh, {``Existence and Stability of Semilocal Strings''},
  Phys.~Rev.~Lett.~{\bf 68},  1263  (1992).

\bibitem{Hindmarsh2}
M.~Hindmarsh, {``Semilocal Topological Defects''},
Nucl.Phys. \textbf{B392},  461 (1993), [arXiv:hep-ph/9206229].

\bibitem{GORS}
G.~W.~Gibbons, M.~E.~Ortiz, F.~Ruiz Ruiz and T.~M.~Samols,
{``Semi-local strings and monopoles''}, Nucl. Phys. \textbf{B385} (1992)
127, [arXiv:hep-th/9203023].

\bibitem{Preskill}
J.~Preskill, {``Semilocal defects''}, Phys.Rev. \textbf{D46}
 (1992) 4218, [arXiv:hep-ph/9206216].

\bibitem{Edelsteinsemilocal}
J.D.~Edelstein,
{``Semi-local Cosmic Strings and the Cosmological Constant Problem''},
Phys.Lett. \textbf{B390} (1997) 101, [arXiv:hep-th/9610163].

\bibitem{ABL99}
A.~Ach\'{u}carro, J.~Borrill and A.~R.~Liddle, ``The formation rate of
semilocal strings,'' Phys.\ Rev.\ Lett.\  {\bf 82}, 3742 (1999)
[arXiv:hep-ph/9802306]; J.~Urrestilla, A.~Ach\'{u}carro, J.~Borrill
and A.~R.~Liddle, ``The evolution and persistence of dumbbells in
electroweak theory,'' JHEP {\bf 0208}, 033 (2002)
[arXiv:hep-ph/0106282].
A.~Ach\'{u}carro and T.~Vachaspati,
``Semilocal and electroweak strings,''
Phys.\ Rept.\  {\bf 327}, 347 (2000)
[arXiv:hep-ph/9904229].

\bibitem{Giddings:2001yu}
S.~B.~Giddings, S.~Kachru and J.~Polchinski, ``Hierarchies from
fluxes in string compactifications,'' Phys.\ Rev.\ D {\bf 66},
106006 (2002) [arXiv:hep-th/0105097].


\bibitem{Felder:2000hj}
G.~N.~Felder, J.~Garcia-Bellido, P.~B.~Greene, L.~Kofman,
A.~D.~Linde and I.~Tkachev, ``Dynamics of symmetry breaking and
tachyonic preheating,'' Phys.\ Rev.\ Lett.\  {\bf 87}, 011601
(2001) [arXiv:hep-ph/0012142];G.~N.~Felder, L.~Kofman and
A.~D.~Linde, ``Tachyonic instability and dynamics of spontaneous
symmetry breaking,'' Phys.\ Rev.\ D {\bf 64}, 123517 (2001)
[arXiv:hep-th/0106179].

\bibitem{pbook}
J.~Polchinski,
``String Theory-II'' Cambridge University Press (1998).

\bibitem{seibergwittenncg}
N.~Seiberg and E.~Witten, ''String Theory and Noncommutative
Geometry,'' JHEP \textbf{9909},032 (1999), [arXiv:hep-th/9908142].




\bibitem{JPP}
C.~V.~Johnson, A.~W.~Peet and J.~Polchinski,
``Gauge theory and the excision of repulson singularities,''
Phys.\ Rev.\ D {\bf 61}, 086001 (2000)
[arXiv:hep-th/9911161].

\bibitem{vafaF}
{C.~Vafa,
``Evidence for F-Theory,''
Nucl.\ Phys.\ B {\bf 469}, 403 (1996),[arXiv:hep-th/9602022].}

\bibitem{drs}
K.~Dasgupta, G.~Rajesh and S.~Sethi,
``M theory, orientifolds and G-flux,''
JHEP {\bf 9908}, 023 (1999), [arXiv:hep-th/9908088];
K.~Becker and K.~Dasgupta,
``Heterotic strings with torsion,''
JHEP {\bf 0211}, 006 (2002)
[arXiv:hep-th/0209077].


\bibitem{senF}
{A.~Sen,
``F-theory and Orientifolds,''
Nucl.\ Phys.\ B {\bf 475}, 562 (1996),[arXiv:hep-th/9605150].}

\bibitem{dm}
{K.~Dasgupta and S.~Mukhi,
``F-theory at constant coupling,''
Phys.\ Lett.\ B {\bf 385}, 125 (1996), [arXiv:hep-th/9606044].}

\bibitem{tate}
J. Tate,
``Algorithm for determining the type of a singular fiber in an
elliptic pencil,'' in {\it Modular Functions of one Variable IV},
Lecture Notes in Maths. Vol. {\bf 476}, Springer-Verlag, Berlin (1975).

\bibitem{witten}
P.~C.~Argyres, M.~Ronen Plesser, N.~Seiberg and E.~Witten,
``New N=2 Superconformal Field Theories in Four Dimensions,''
Nucl.\ Phys.\ B {\bf 461}, 71 (1996)
[arXiv:hep-th/9511154].

\bibitem{ad}
{P.~C.~Argyres and M.~R.~Douglas,
``New phenomena in SU(3) supersymmetric gauge theory,''
Nucl.\ Phys.\ B {\bf 448}, 93 (1995), [arXiv:hep-th/9505062].}

\bibitem{GZ}
M.~R.~Gaberdiel and B.~Zwiebach,
``Exceptional groups from open strings,''
Nucl.\ Phys.\ B {\bf 518}, 151 (1998)
[arXiv:hep-th/9709013].



\bibitem{HHKV}
M.~Hindmarsh, R.~Holman, T.~W.~Kephart and T.~Vachaspati,
``Generalized semilocal theories and higher Hopf maps,''
Nucl.\ Phys.\ B {\bf 404}, 794 (1993)
[arXiv:hep-th/9209088].


\bibitem{sw}
N.~Seiberg and E.~Witten,
``Electric - magnetic duality, monopole condensation, and confinement in N=2
supersymmetric Yang-Mills theory,''
Nucl.\ Phys.\ B {\bf 426}, 19 (1994)
[Erratum-ibid.\ B {\bf 430}, 485 (1994)], [arXiv:hep-th/9407087].

\bibitem{aharony}
O.~Aharony, J.~Sonnenschein, S.~Yankielowicz and S.~Theisen,
``Field theory questions for string theory answers,''
Nucl.\ Phys.\ B {\bf 493}, 177 (1997), [arXiv:hep-th/9611222];
{M.~R.~Douglas, D.~A.~Lowe and J.~H.~Schwarz,
``Probing F-theory with multiple branes,''
Phys.\ Lett.\ B {\bf 394}, 297 (1997), [arXiv:hep-th/9612062].}

\bibitem{gsvy}
{B.~R.~Greene, A.~D.~Shapere, C.~Vafa and S.~T.~Yau,
``Stringy Cosmic Strings And Noncompact Calabi-Yau Manifolds,''
Nucl.\ Phys.\ B {\bf 337}, 1 (1990).}

\bibitem{shiu}
{B.~R.~Greene, K.~Schalm and G.~Shiu,
``Warped compactifications in M and F theory,''
Nucl.\ Phys.\ B {\bf 584}, 480 (2000), [arXiv:hep-th/0004103].}





\bibitem{KKLV}  R.~Kallosh, L.~Kofman, A.~Linde and A.~Van Proeyen,
{``Superconformal Symmetry, Supergravity and Cosmology''},
 Class.Quant.Grav. \textbf{17} (2000) 4269,
 [arXiv:hep-th/0006179].






 \bibitem{PRTT96}
A.~A.~Penin, V.~A.~Rubakov, P.~G.~Tinyakov and S.~V.~Troitsky,
``What becomes of vortices in theories with flat directions,''
Phys.\ Lett.\ B {\bf 389}, 13 (1996)
[arXiv:hep-ph/9609257].



\bibitem{ADPU02}
A.~Ach\'{u}carro, A.~C.~Davis, M.~Pickles and J.~Urrestilla,
``Vortices in theories with flat directions,''
Phys.\ Rev.\ D {\bf 66}, 105013 (2002)
[arXiv:hep-th/0109097];
A.~Ach\'{u}carro, A.~C.~Davis, M.~Pickles and J.~Urrestilla,
``Fermion zero modes in N = 2 supervortices,''
Phys.\ Rev.\ D {\bf 68}, 065006 (2003)
[arXiv:hep-th/0212125].



\bibitem{Pickles:2002ym}
M.~Pickles and J.~Urrestilla, ``Nielsen-Olesen strings in supersymmetric
models,'' JHEP {\bf 0301}, 052 (2003) [arXiv:hep-th/0211240].







\bibitem{lindebook} A.D. Linde,  {\it Particle  Physics  and
Inflationary Cosmology,} (Harwood, Chur, Switzerland, 1990).



\bibitem{VilenkinShellard}
A. Vilenkin and E.P.S. Shellard, {\it Cosmic strings and other
topological defects,} Cambridge Univ. Press (Cambridge 1994).

\bibitem{Tkachev:1998dc}
I.~Tkachev, S.~Khlebnikov, L.~Kofman and A.~D.~Linde,
``Cosmic strings from preheating,''
Phys.\ Lett.\ B {\bf 440}, 262 (1998)
[arXiv:hep-ph/9805209].

\bibitem{Bastero-Gil:1999fz}
M.~Bastero-Gil, S.~F.~King and J.~Sanderson, ``Preheating in
supersymmetric hybrid inflation,'' Phys.\ Rev.\ D {\bf 60}, 103517
(1999) [arXiv:hep-ph/9904315].





















\end{thebibliography}
\end{document}